# About some properties of the canonical density matrix versus the canonical Bloch equation

**Dušan POPOV**

University Politehnica Timisoara, Romania
Department of Physical Foundations of Engineering
Member of Serbian Academy of Nonlinear Sciences (SANS), Belgrade, Serbia
E-mail: dusan_popov@yahoo.co.uk
ORCID: https://orcid.org/0000-0003-3631-3247

## Abstract

We examine some properties of the non-normalized (or canonical) density matrix in the coherent states representation, by two equivalent ways. On the one hand by its definition, and on the other hand as a solution to Bloch's canonical equation. It is concluded that, since in many cases Bloch's differential equation is difficult to solve, in applications it is preferable to use the expression obtained directly from the definition of the canonical density matrix in the coherent states representation. This conclusion is verified by examining several cases of quantum systems with linear or quadratic energy spectrum.

## 1. Introduction

As it is known from quantum mechanics books, the states of a quantum system can be of two types: pure states, whose description is possible through a state vector $|\Psi_n>$ or, in some representation $X \equiv \{x\}$, through a wave function $\Psi_n(x) \equiv <\Psi_n | x>$, respectively mixed states, whose description requires the use of a statistical operator - the density operator $\rho$ or, in some representation, a density matrix $<x|\rho|x'>$. In fact, the density operator is a linear normalized Hermitian operator, with trace unity. For a pure state, it is the outer product of a state with itself, $\rho_n = |\Psi_n><\Psi_n|$. For a statistical ensemble or a statistical mixture we have insufficient information about the system, so we must perform a statistical average in order to describe quantum observables. In a statistical ensemble, which is a mixture of pure states, each participating with a certain probability $w_n$ , the density operator is a sum over all pure states each with corresponding probabilities:

$$\rho = \sum_n w_n | \Psi_n >< \Psi_n | \tag{1.01}$$

In some arbitrary representation $X \equiv \{x\}$, the corresponding density matrix looks like this

$$< x | \rho | x' > = \sum_n w_n \, \Psi_n^*(x) \Psi_n(x') \tag{1.02}$$

Generally speaking, using the density matrix approach, the calculations are greatly simplified.

One of the most frequently encountered statistical ensembles is the *canonical ensemble*, that represents the possible quantum states in thermal equilibrium with a heat bath at a fixed temperature $T = (k_B \beta)^{-1}$, where $k_B$ is the Boltzmann's constant and $\beta$ is the inverse temperature parameter.

For a quantum system with the Hamiltonian $\hat{\mathcal{H}}$ the eigenvalue equation is

$$\hat{\mathcal{H}}(x)\Psi_n(x) = E(n)\Psi_n(x) = \hbar\omega\, e(n)\Psi_n(x) \tag{1.03}$$

In some representation, e.g. in the coordinate representation $\{x\}$, the density matrix corresponding to a canonical ensemble may be written in general form as

$$< x | \hat{\rho}(\beta) | x' > \equiv \rho(x, x'; \beta) = \frac{1}{Z(\beta)} \sum_n e^{-\beta E(n)} \, \Psi_n(x) \Psi_n^*(x') \tag{1.04}$$

where the partition function is (we suppose that the energy levels are non degenerated)

$$Z(\beta) = \sum_n e^{-\beta E(n)} \tag{1.05}$$

If the quantum system obeys the canonical distribution (that is, the system only exchanges energy with the environment it is in contact with and does not exchange particles) the normalized density operator is

$$\hat{\rho}(\beta) = \frac{1}{Z(\beta)} \exp\left(-\beta \hat{\mathcal{H}}\right) \equiv \frac{1}{Z(\beta)} \hat{\Omega}(\beta) \quad , \quad \hat{\Omega}(\beta) = \exp\left(-\beta \hat{\mathcal{H}}\right) \tag{1.06}$$

Consequently, $\Omega(\beta)$ is the *non – normalized canonical density operator* or *canonical operator*. If we regard the density operator as function of the temperature parameter $\beta = (k_B T)^{-1}$, then for extremely high temperatures, the density operator gets to the identity operator. This may be regarded as an "initial" or boundary condition for density operator:

$$\lim_{\beta \to 0} \hat{\Omega}(\beta) \equiv \hat{\Omega}(0) = I \tag{1.07}$$

Using the energy eigenequation as well as the completeness relation for Fock vectors

$$\sum_n | n >< n | = 1 \tag{1.08}$$

we can write the density operator in the following manner

$$\hat{\Omega}(\beta) = \sum_n \exp[-\beta \hbar \omega \, e(n)] \, | n >< n | \tag{1.09}$$

As function of the temperature parameter $\beta$, the density operator will satisfy the so – called *canonical Bloch equation*, which was formulated for the first time by Felix Bloch [1]. Later, the various aspects of Bloch's equation have been examined in several papers [2], [3], [4], [5], [6], [7], [8]:

$$-\frac{\partial}{\partial \beta} \hat{\Omega}(\beta) = \hat{\mathcal{H}} \, \hat{\Omega}(\beta) , \qquad \lim_{\beta \to 0} \hat{\Omega}(\beta) \equiv \hat{\Omega}(0) = \hat{I} \tag{1.10}$$

In the coordinate representation the non-normalized density matrix is

$$< x | \hat{\Omega}(\beta) | x' > \equiv \Omega(x, x'; \beta) = \sum_n \exp[-\beta \hbar \omega \, e(n)] \, \Psi_n(x) \Psi_n^*(x') \tag{1.11}$$

Analogously, the Bloch equation for the density matrix in the coordinate representation is

$$-\frac{\partial}{\partial \beta} \Omega(x, x'; \beta) = \hat{\mathcal{H}}(x) \Omega(x, x'; \beta) \tag{1.12}$$

where the Hamiltonian operator $\mathcal{H}(x)$ acts only on the variable from the left side, i.e. $x$.

The boundary condition for density matrix is then expressed through Dirac delta distribution

$$\lim_{\beta \to 0} \Omega(x, x'; \beta) \equiv \sum_n \Psi_n(x) \Psi_n^*(x') = \delta(x - x') \tag{1.13}$$

Because the one-dimensional problems the Hamiltonian operator is

$$\hat{\mathcal{H}}(x) = -\frac{\hbar^2}{2m} \frac{\partial^2}{\partial x^2} + V(x) \tag{1.14}$$

it is observed that that the Bloch equation is an differential equation with partial derivatives and initial condition. Unfortunately, there are only few quantum systems for which Bloch equation can be solved exactly, i.e. systems for which it can find expression for the density matrix $\Omega(x, x'; \beta)$ [2].

The derivation of Bloch's equation in the coordinate representation is simple. We will successively have:

$$\begin{aligned} -\frac{\hbar^2}{2m} \frac{\partial^2}{\partial x^2} \Omega(x, x'; \beta) &= \sum_n \exp[-\beta \hbar \omega \, e(n)] \left[ -\frac{\hbar^2}{2m} \frac{\partial^2}{\partial x^2} \Psi_n(x) \right] \Psi_n^*(x') = \\ = \sum_n \exp[-\beta \hbar \omega \, e(n)] \, [\mathcal{H}(x) - V(x)] \Psi_n(x) \Psi_n^*(x') &= -\frac{\partial}{\partial \beta} \Omega(x, x'; \beta) - V(x) \Omega(x, x'; \beta) \end{aligned} \tag{1.15}$$

On the other hand, the calculations related to the density operator in the representation of coherent states proved to be very suitable and efficient.

**2. Coherent states representation**

For the first time, the coherent states (CSs) were introduced in 1926 by Schrödinger [9], as states that has dynamics most closely resembling the oscillatory behavior of a classical one dimensional harmonic oscillator (HO-1D). After several decades of "silence", in which it seemed that the concept of coherent states is not too exciting, it began to emerge a stream of articles and books in which have been proposed different kinds of CSs, referring not only on the HO-1D but also on other quantum systems. In this sense, an excellent review of different kinds of CSs is made in the Dodonov’s paper [10].

Generally, a coherent state (CS) can be expressed as a ket vector labeled by a complex number $z = |z| \exp(i\varphi)$, with $|z| \leq \mathcal{R}_c \leq \infty$ , $0 \leq \varphi \leq 2\pi$, where $\mathcal{R}_c$ is the convergence radius. Their expansion in the Fock – vectors basis $\{\, |n>,\ n = 0,1,2,...,n_{\max} \leq \infty \}$ is

$$|z> = \frac{1}{\sqrt{\mathcal{N}(|z|^2)}} \sum_n \frac{z^n}{\sqrt{\rho(n)}} |n> \tag{2.01}$$

Here $\rho(n)$ are positive numbers which essentially determine the internal structure of CSs, and for this reason $\rho(n)$ are called the *structure constants*.

The CSs exists if and only if the normalization function $\mathcal{N}(|z|^2)$ is expressed through an analytical function of real variable $|z|^2$. In addition, any CSs must fulfill a number of criteria summarized by Klauder (often called "*Klauder's prescriptions*"): continuity in complex label, normalization, non-orthogonality, resolution of unity operator with unique positive weight function of the integration measure, and supplementary, only for the CSs corresponding to the systems with infinite energy spectrum, the temporal stability and action identity [11].

On the one hand, if $\rho(n) = n!$, the CSs are called *canonical* or *linear* CSs which are characteristic for HO-1D, whose creation (raising) $a^+$ and annihilation (lowering) $a$ operators are canonically conjugated: $\left[a, a^+\right] = 1$. On the other hand, if $\rho(n) = n!\, f(n) \neq n!$, where $f(n)$ is the nonlinearity function, we have to deal with different kinds of *nonlinear* CSs, depending on the structure constants.

But the most general class of CSs is the so called *generalized hypergeometric coherent states* (GHG-CSs) whose appellation becomes from their normalization function which is given by a generalized hypergeometric function, e.g. ${}_pF_q\left(\{a_i\}_{i=1}^p;\ \{b_j\}_{j=1}^q;\ |z|^2\right)$. These kinds of states were firstly introduced by Appl and Schiller [12] and applied to the mixed (thermal) states in one of our previous papers [13]. Their expansion in the Fock-vectors basis is

$$|z> = \frac{1}{\sqrt{{}_pF_q\left(\boldsymbol{a};\, \boldsymbol{b};\, |z|^2\right)}} \sum_{n=0}^{\infty} \frac{z^n}{\sqrt{\rho(n)}} |n> \tag{2.02}$$

Here and in the next, we will use the short notation for the sequence of real numbers $\left\{a_1,\, a_2, ..., a_p\right\} \equiv \{a_i\}_{i=1}^p \equiv \boldsymbol{a}$ and so on. By particularizing in a suitable manner these numbers, as well as the positive integers $p$ and $q$, we can obtain all known CSs which have the physical meaning [14].

In the next, for the sake of simplicity of calculations we shall work with the *non-normalized* or *canonical* CSs , $|z))$ , i.e.

$$|z>=\frac{1}{\sqrt{{}_pF_q\left(\mathbf{a};\,\mathbf{b};\,|z|^2\right)}}\,|z))\quad ,\qquad |z))\equiv\sum_{n=0}^{\infty}\frac{z^n}{\sqrt{\rho(n)}}|n> \tag{2.03}$$

The generalized hypergeometric function is defined as [15]

$${}_pF_q\left(\mathbf{a};\,\mathbf{b};\varsigma\right)=\sum_{n=0}^{\infty}\frac{\prod_{i=1}^{p}(a_i)_n}{\prod_{j=1}^{q}(b_j)_n}\frac{\varsigma^n}{n!}\equiv\sum_{n=0}^{\infty}\frac{1}{\rho(n)}\varsigma^n \tag{2.04}$$

by using the Euler gamma functions $\Gamma(a)$ and the Pochhammer symbols $(a)_n=\Gamma(a+n)/\Gamma(a)$ .

The structure constants are defined as

$$\rho(n)=n!\frac{\prod_{j=1}^{q}(b_j)_n}{\prod_{i=1}^{p}(a_i)_n}=\frac{\prod_{i=1}^{p}\Gamma(a_i)}{\prod_{j=1}^{q}\Gamma(b_j)}\,\frac{\Gamma(n+1)\prod_{j=1}^{q}\Gamma(b_j+n)}{\prod_{i=1}^{p}\Gamma(a_i+n)} \tag{2.05}$$

It is useful also to remember relation between the generalized hypergeometric function ${}_pF_q\left(\ldots;\varsigma\right)$ and Meijer's functions $G_{p,q}^{m,n}\left(\varsigma\middle|\ldots\right)$ [15].

$${}_pF_q\left(\mathbf{a};\,\mathbf{b};\,\varsigma\right)=\Gamma_{qp}(b/a)\,G_{p,q+1}^{1,p}\left(-\varsigma\,\middle|\,\begin{matrix}\mathbf{1-a}\;; & /\\ 0\;\;; & \mathbf{1-b}\end{matrix}\right) \tag{2.06}$$

As well as the normalized GHG-CSs $|z>$, the non-normalized GHG-CSs $|z))$ meet all Klauder's criteria imposed to a coherent state [11], but with some differences that relate to properties of temporal stability and action identity. These differences are related to the character of the energy spectrum: linear or non-linear in relation to the main quantum number $n$. So, the non-normalized CSs $|z))$, defined as

$$|z))\equiv\sum_{n=0}^{\infty}\frac{z^n}{\sqrt{\rho(n)}}|n> \tag{2.07}$$

satisfy the following properties: *a*.) continuity in the complex label $z$ , i.e. if $z'\to z$, then $|z'))\to|z))$ , *b*.) the non-orthogonality

$$((z\,|\,z'))=\sum_{n=0}^{\infty}\frac{\left(z^*z'\right)^n}{\rho(n)}={}_pF_q\left(\mathbf{a};\,\mathbf{b};\,z^*z'\right)\neq 0 \tag{2.08}$$

and *c*.) the unity operator decomposition

$$\int d\tilde{\mu}(z) \mid z))((z \mid = \sum_n \mid n >< n \mid = 1 \tag{2.09}$$

Now, let us we evaluate the positive integration measure $d\tilde{\mu}(z) = \frac{d\varphi}{2\pi} d\left(\mid z \mid^2\right) \tilde{h}\left(\mid z \mid^2\right)$ or, more precisely, their positive defined weight function $\tilde{h}\left(\mid z \mid^2\right)$.

$$1 = \int d\tilde{\mu}(z) \mid z))((z \mid = \int \frac{d\varphi}{2\pi} d\left(\mid z \mid^2\right) \tilde{h}\left(\mid z \mid^2\right) \sum_{n,m} \frac{\left(z^*\right)^n}{\sqrt{\rho(n)}} \mid n >< m \mid \frac{(z)^m}{\sqrt{\rho(m)}} \tag{2.10}$$

The angular integral is

$$\int_0^{2\pi} \frac{d\varphi}{2\pi} \left(z^*\right)^n (z)^m = \left(\mid z \mid^2\right)^n \delta_{nm} \tag{2.11}$$

Using the completeness relation of the Fock vectors, we obtain

$$1 = \sum_n \frac{\mid n >< n \mid}{\rho(n)} \int_0^{\mathcal{R}_c} d\left(\mid z \mid^2\right) \tilde{h}\left(\mid z \mid^2\right) \left(\mid z \mid^2\right)^n = \sum_n \mid n >< n \mid \tag{2.12}$$

from which it is evident that we must have

$$\int_0^{\mathcal{R}_c} d\left(\mid z \mid^2\right) \tilde{h}\left(\mid z \mid^2\right) \left(\mid z \mid^2\right)^n = \rho(n) = \frac{\prod_{i=1}^{p} \Gamma(a_i)}{\prod_{j=1}^{q} \Gamma(b_j)} \; \frac{\Gamma(n+1) \prod_{j=1}^{q} \Gamma(b_j + n)}{\prod_{i=1}^{p} \Gamma(a_i + n)} \quad . \tag{2.13}$$

The convergence radius of the CSs $\mathcal{R}_c$ is the convergence radius of the hypergeometric series, given by one of the convergence criteria of the power series [14] :

$$\mathcal{R}_c = \lim_{n \to \infty} \inf \sqrt[n]{\rho(n)} = \lim_{n \to \infty} \frac{\rho(n)}{\rho(n+1)} = \begin{cases} \infty & \text{Stieltjes moment problem} \\ < \infty & \text{Hausdorff moment problem} \end{cases} . \tag{2.14}$$

Depending on the value of the convergence radius $\mathcal{R}_c$ this relation is the Haussdorff moment problem, if $\mathcal{R}_c < \infty$, respectively the Stieltjes moment problem if $\mathcal{R}_c = \infty$.

The standard method to find the weight function of the integration measure is to solve the above integral with the exponent change $n = s - 1$, and then to obtain

$$\int_0^{\mathcal{R}_c} d\left(|z|^2\right)\tilde{h}\left(|z|^2\right)\left(|z|^2\right)^{s-1} = \frac{\prod_{i=1}^{p}\Gamma(a_i)}{\prod_{j=1}^{q}\Gamma(b_j)} \frac{\Gamma(s)\prod_{j=1}^{q}\Gamma(b_j-1+s)}{\prod_{i=1}^{p}\Gamma(a_i-1+s)} \quad . \tag{2.15}$$

Generally, their solution, i.e. the weight function of the integration measure $\tilde{h}\left(|z|^2\right)$ is expressed through the Meijer's $G$-functions [15]:

$$\tilde{h}\left(|z|^2\right) = \Gamma_{pq}(a/b)\; G_{p,q+1}^{q+1,0}\left(|z|^2 \left| \begin{array}{ll} /; & \boldsymbol{a-1} \\ 0, \boldsymbol{b-1}\,; & / \end{array}\right.\right) . \tag{2.16}$$

so that the integration measure $d\tilde{\mu}(z)$ is

$$\begin{aligned} d\tilde{\mu}(z) &= \Gamma_{pq}(a/b)\,\frac{d^2 z}{\pi}\, G_{p,q+1}^{q+1,0}\left(|z|^2 \left| \begin{array}{ll} / \;\; ; & \boldsymbol{a-1} \\ 0,\;\; \boldsymbol{b-1}\;\; ; & / \end{array}\right.\right) \equiv \\ &\equiv \frac{\prod_{i=1}^{p}\Gamma(a_i)}{\prod_{j=1}^{q}\Gamma(b_j)}\, d\left(|z|^2\right)\frac{d\varphi}{2\pi}\, G_{p,q+1}^{q+1,0}\left(|z|^2 \,|\, \ldots\right) \end{aligned} \tag{2.17}$$

This expression must ensure the unity operator decomposition. In this regard we have, successively

$$\begin{aligned} \int d\tilde{\mu}(z)\,|\,z))((z\,| &= \Gamma_{pq}(a/b)\sum_{n,n'}\frac{|\,n><n'|}{\sqrt{\rho(n)\,\rho(n')}}\int_0^{\mathcal{R}_c} d\left(|z|^2\right) G_{p,q+1}^{q+1,0}\left(|z|^2\,|\ldots\right)\int_0^{2\pi}\frac{d\varphi}{2\pi}\left(z^*\right)^n z^{n'} = \\ &= \Gamma_{pq}(a/b)\sum_n \frac{|\,n><n\,|}{\rho(n)}\int_0^{\mathcal{R}_c} d\left(|z|^2\right)\left(|z|^2\right)^n G_{p,q+1}^{q+1,0}\left(|z|^2\,|\ldots\right) = \sum_n |\,n><n\,| = 1 \end{aligned} \tag{2.18}$$

Consequently, it is useful to remember the following integral that will appear several times further, in which both types of moments (Stieltjes and Hausdorff) are expressed by a single integral

$$\begin{aligned} &\int_0^{\mathcal{R}_c\to\infty} d\left(|z|^2\right)\left(|z|^2\right)^n H\left(\mathcal{R}_c - |z|^2\right) G_{p,q+1}^{q+1,0}\left(|z|^2 \left| \begin{array}{ll} / \;\; ; & \boldsymbol{a-1} \\ 0,\;\; \boldsymbol{a-1}\;\; ; & / \end{array}\right.\right) = \\ &= \frac{\prod_{i=1}^{p}\Gamma(a_i)}{\prod_{j=1}^{q}\Gamma(b_j)}\,\rho(n) = \Gamma(n+1)\frac{\prod_{j=1}^{q}\Gamma(b_j+n)}{\prod_{i=1}^{p}\Gamma(a_i+n)} \end{aligned} \tag{2.19}$$

Here we used the Heaviside step function

$$H\left(\mathcal{R}_c - |z|^2\right) = \begin{cases} 0, & |z|^2 > \mathcal{R}_c \\ 1, & |z|^2 \le \mathcal{R}_c \end{cases} . \tag{2.20}$$

### 3. The Bloch equation in the coherent states representation

To begin with, let's remind ourselves of some mathematical properties of the operator $x\frac{\partial}{\partial x}$. By the way, the operator $x\frac{\partial}{\partial x}$ is widely used in the case of power series (see, for example [16]). In calculations, will be useful to evince the following actions of a function $\hat{f}\left(x\frac{\partial}{\partial x}\right)$ of operators $x$:

$$\begin{gathered} \hat{f}\left(x\frac{\partial}{\partial x}\right)(\pm x)^n = f(n)\,(\pm x)^n \\ \hat{f}\left(x\frac{\partial}{\partial x}\right)g(x) = \hat{f}\left(x\frac{\partial}{\partial x}\right)\left(\sum_{n=0}^{\infty} g_n x^n\right) = \sum_{n=0}^{\infty} g_n\ f(n)x^n \\ \hat{f}\left(x\frac{\partial}{\partial x}\right)\exp\left[-\beta\,\hat{f}\left(x\frac{\partial}{\partial x}\right)\right] = \exp\left[-\beta\,\hat{f}\left(x\frac{\partial}{\partial x}\right)\right]\hat{f}\left(x\frac{\partial}{\partial x}\right) \end{gathered} \tag{3.01}$$

Also, some properties of Pochhammer's symbol

$$(x)_n = \frac{\Gamma(x+n)}{\Gamma(x)} = (x+n-1)(x+n-2)...(x+1)x \tag{3.02}$$

$$\frac{(x)_n}{(x+n-1)} = (x)_{n-1} \tag{3.03}$$

$$(x)_{n+1} = (x+n)(x)_n \qquad , \qquad (x+1)_n = \frac{1}{x}(x+n)(x)_n = \frac{1}{x}(x)_{n+1} \tag{3.04}$$

The most general coherent states are the so-called *generalized hypergeometrical coherent states* (GH-CSs). They are generated by a pair of Hermitical creation $\hat{\mathcal{A}}_+$ and annihilation $\hat{\mathcal{A}}_-$ operators, whose actions on the Fock vectors are [17]

$$\begin{gathered} \hat{\mathcal{A}}_- \,|\, n> = \sqrt{e(n)}\,|\, n-1> \,, \qquad <n\,|\,\hat{\mathcal{A}}_+ = \sqrt{e(n+1)} < n+1\,| \\ <n\,|\,\hat{\mathcal{A}}_+\hat{\mathcal{A}}_-\,|\,n> = e(n) \equiv <n\,|\,\hat{\mathcal{H}}\,|\,n> \end{gathered} \tag{3.05}$$

where $e(n)$ are the (dimensionless) eigenvalues of the Hamiltonian's operator.

The actions of the creation $\hat{\mathcal{A}}_+$ and annihilation $\hat{\mathcal{A}}_-$ operators on the vacuum state (state without particles) $|\,0>$ are

$$|n> = \frac{1}{\sqrt{\rho(n)}}\left(\hat{\mathcal{A}}_+\right)^n |0> \quad , \qquad <n| = \frac{1}{\sqrt{\rho(n)}} <0|\left(\hat{\mathcal{A}}_-\right)^n \tag{3.06}$$

With these considerations, the non-normalized coherent states $|z))$ (and their adjuncts) $((z^*|$ can be expressed as a hypergeometric operational function

$$|z)) = \sum_n \frac{\left(z\,\hat{\mathcal{A}}_+\right)^n}{\rho(n)}|0> \quad , \qquad ((z^*| = <0|\sum_n \frac{\left(z^*\,\hat{\mathcal{A}}_-\right)^n}{\rho(n)} \tag{3.07}$$

The statement that these are the most general coherent states is based on the assertion that, by particularizing the integer indices $p$ and $q$, as well as the expression of the eigenvalues of the energy $e(n)$, all the known coherent states are obtained.

For now let's recall the commutation relation between the canonical annihilation $\hat{a}$ and creation $\hat{a}^+$ operators, as well as the particle number operator $\hat{\mathcal{N}}$ :

$$\begin{gathered} \hat{a}|n> = \sqrt{n}\,|n-1> , \;\; \hat{a}^+|n> = \sqrt{n+1}\,|n+1> \quad , \quad [\hat{a}\,,\hat{a}^+] = \hat{a}\hat{a}^+ - \hat{a}^+\hat{a} = 1 \;, \\ <n|\hat{a}^+\hat{a}|n> = n \quad , \quad \hat{\mathcal{N}} = \hat{a}^+\hat{a} \quad , \quad \hat{\mathcal{N}}|n> = n|n> \end{gathered} \tag{3.08}$$

In the Bargmann space, the algebra of ladder operators $\hat{a}^+$ and $\hat{a}$ is represented directly by the algebra of the complex variable $z$ and its derivative $\frac{\partial}{\partial z}$ , respectively the particle number $n$ by the product $\hat{a}^+ a = z\frac{\partial}{\partial z}$ [18]. If we adopt this representation of the above operators, in the Segal-Bargmann complex space [19], we will obtain:

$$a = \frac{\partial}{\partial z} \quad , \quad \hat{a}^+ = z \quad , \quad [\hat{a}\,,\hat{a}^+] = \frac{\partial}{\partial z}\left(z\,\circ\right) - z\frac{\partial}{\partial z} = 1 \;, \qquad \hat{\mathcal{N}} = z\frac{\partial}{\partial z} \tag{3.09}$$

Their actions on the non-normalized coherent states is

$$a|z)) = z|z)) \;\; , \;\; ((z^*|\hat{a}^+ = z^*((z^*| \;\; , \quad ((z^*|\hat{a}^+ a|z)) = ((z^*|\hat{\mathcal{N}}|z)) = |z|^2 \tag{3.10}$$

Now, let us consider a certain operator $\hat{O}$, which characterize the studied quantum system. Its matrix elements in the representation of normalized GHG-CSs are

$$\begin{aligned} <z^*|\hat{O}|z'> &\equiv \frac{1}{\sqrt{{}_pF_q\left(\boldsymbol{a};\boldsymbol{b};|z|^2\right)}}\frac{1}{\sqrt{{}_pF_q\left(\boldsymbol{a};\boldsymbol{b};|z'|^2\right)}}\sum_{n,n'=0}^{\infty}\frac{\left(z^*\right)^n\left(z'\right)^{n'}}{\sqrt{\rho(n)}\sqrt{\rho(n')}}<n|\hat{O}|n'> = \\ &= \frac{1}{\sqrt{{}_pF_q\left(\boldsymbol{a};\boldsymbol{b};|z|^2\right)}}\frac{1}{\sqrt{{}_pF_q\left(\boldsymbol{a};\boldsymbol{b};|z'|^2\right)}}((z^*|\hat{O}|z')) \end{aligned} \tag{3.11}$$

where $((z^*|\hat{O}|z'))$ are their non-normalized matrix elements in representation of coherent states. If the operator $\hat{O}$ is diagonal in the base of the Fock vectors, i.e. their eigenvalue equation is $\hat{O}|n> = O(n)|n>$, then the expression is simplified

$$((z^*|\hat{O}|z')) \equiv \sum_{n=0}^{\infty} O(n)\frac{\left(z^* z'\right)^n}{\rho(n)} \tag{3.12}$$

Generally, the eigenvalues can be expanded in a power series of the main quantum number $n$:

$$O(n) = \sum_{m=0}^{\infty} c_m \, n^m \tag{3.13}$$

If we consider the operational equality

$$n^m \left(z^* z'\right)^n = \left( z^* \frac{\partial}{\partial z^*} \right)^m \left(z^* z'\right)^n \tag{3.14}$$

then the matrix elements can be written as

$$\begin{aligned} ((z^* \,|\, \hat{O} \,|\, z')) &\equiv \sum_{n=0}^{\infty} O(n) \frac{\left(z^* z'\right)^n}{\rho(n)} = \sum_{n=0}^{\infty} \left[ \sum_{m=0}^{\infty} c_m \left( z^* \frac{\partial}{\partial z^*} \right)^m \right] \frac{\left(z^* z'\right)^n}{\rho(n)} = \\ &= \sum_{n=0}^{\infty} \hat{O}\left( z^* \frac{\partial}{\partial z^*} \right) \frac{\left(z^* z'\right)^n}{\rho(n)} = \hat{O}\left( z^* \frac{\partial}{\partial z^*} \right) \sum_{n=0}^{\infty} \frac{\left(z^* z'\right)^n}{\rho(n)} = \hat{O}\left( z^* \frac{\partial}{\partial z^*} \right) {}_pF_q\left(\boldsymbol{a};\, \boldsymbol{b};\, z^* z'\right) \end{aligned} \tag{3.15}$$

This means that any matrix element in the representation of coherent states can be written in the form of an operator depending on the product $z^* \frac{\partial}{\partial z^*}$ which acts on the generalized hypergeometric series ${}_pF_q\left(\boldsymbol{a};\, \boldsymbol{b};\, z^* z'\right)$, having the argument $z^* z'$.

This is the *fundamental property* on which the calculations in this paper are based.

Let's evaluate this idea for some physical observables that we will need further.

$$\hat{\mathcal{H}} |n\rangle = E(n) |n\rangle = \hbar \omega e(n) |n\rangle \tag{3.16}$$

$$((z^* \,|\, \hat{\mathcal{H}} \,|\, z')) = \hat{\mathcal{H}}\left( z^* \frac{\partial}{\partial z^*} \right) \sum_{n=0}^{\infty} \frac{\left(z^* z'\right)^n}{\rho(n)} = \hbar \omega e\left( z^* \frac{\partial}{\partial z^*} \right) {}_pF_q\left(\boldsymbol{a};\, \boldsymbol{b};\, z^* z'\right) \tag{3.17}$$

From the definition of the non-normalized density operator it is obtained

$$((z^* \,|\, \hat{\Omega} \,|\, z')) \equiv ((z^* \,|\, e^{-\beta \hat{\mathcal{H}}} \,|\, z')) = \sum_{n=0}^{\infty} e^{-\beta \hbar \omega e(n)} \frac{\left(z^* z'\right)^n}{\rho(n)} = e^{-\beta \hbar \omega e\left( z^* \frac{\partial}{\partial z^*} \right)} {}_pF_q\left(\boldsymbol{a};\, \boldsymbol{b};\, z^* z'\right) \tag{3.18}$$

$$\begin{aligned} ((z^* \,|\, \hat{\mathcal{H}} \hat{\Omega} \,|\, z')) &= ((z^* \,|\, \hat{\mathcal{H}} \left[ \int d\mu(\sigma) |\, \sigma))((\sigma^* \,| \right] \hat{\Omega} \,|\, z')) = \\ &= \int d\mu(\sigma) ((z^* \,|\, \hat{\mathcal{H}} \,|\, \sigma))((\sigma^* \,|\, \hat{\Omega} \,|\, z')) = \int d\mu(\sigma) \hbar \omega e\left( z^* \frac{\partial}{\partial z^*} \right) ((z^* \,|\, \sigma))((\sigma^* \,|\, \hat{\Omega} \,|\, z')) = \\ &= \hbar \omega e\left( z^* \frac{\partial}{\partial z^*} \right) ((z^* \,| \left[ \int d\mu(\sigma) |\, \sigma))((\sigma^* \,| \right] |\, \hat{\Omega} \,|\, z')) = \hbar \omega e\left( z^* \frac{\partial}{\partial z^*} \right) ((z^* \,|\, \hat{\Omega} \,|\, z')) \end{aligned} \tag{3.19}$$

Taking into account the above considerations, the corresponding *canonical Bloch equation* in the coherent states representation, in the most general form

$$-\frac{\partial}{\partial\beta}((z^* \,|\, \hat{\Omega} \,|\, z')) = ((z^* \,|\, \hat{\mathcal{H}}\hat{\Omega} \,|\, z')) \tag{3.20}$$

then becomes

$$-\frac{\partial}{\partial\beta}((z^* \,|\, \hat{\Omega} \,|\, z')) = \hat{\mathcal{H}}\left( z^* \frac{\partial}{\partial z^*} \right)((z^* \,|\, \hat{\Omega} \,|\, z')) \tag{3.21}$$

This means that in the right-hand side of Bloch equation, in the expression of Hamilton's operator $\hat{\mathcal{H}}$ , each quantum number $n$ in the energy eigenvalues $e(n)$ must be replaced by the operator $z^* \frac{\partial}{\partial z^*}$ , i.e. we obtain $e\left( z^* \frac{\partial}{\partial z^*} \right)$:

$$-\frac{\partial}{\partial\beta}((z^* \,|\, \hat{\Omega} \,|\, z')) = \hbar\omega e\left( z^* \frac{\partial}{\partial z^*} \right)((z^* \,|\, \hat{\Omega} \,|\, z')) \tag{3.22}$$

This result can be verified by the direct derivation of expression (3.18) of the non-normalized density matrix, obtained in the coherent states representation:

$$\begin{aligned} -\frac{\partial}{\partial\beta}((z^* \,|\, \hat{\Omega} \,|\, z')) &= -\frac{\partial}{\partial\beta}\sum_{n=0}^{\infty} e^{-\beta\hbar\omega e(n)} \frac{\left(z^* z'\right)^n}{\rho(n)} = \\ = \hbar\omega\sum_{n=0}^{\infty} e(n) e^{-\beta\hbar\omega e(n)} \frac{\left(z^* z'\right)^n}{\rho(n)} &= \hbar\omega\, e\left( z^* \frac{\partial}{\partial z^*} \right)((z^* \,|\, \hat{\Omega} \,|\, z')) \end{aligned} \tag{3.23}$$

The same result, but using a slightly different reasoning, was obtained by [5], for a particular case of coherent states, namely the canonical coherent states, those of the one-dimensional harmonic oscillator (HO-1D). Fortunately, our calculation above can be applied to *any pair* of Hermitian operators, respectively GHG-CSs.

In conclusion, in the expression of $e(n)$ we did the replacement $n \to z^* \frac{\partial}{\partial z^*}$.

In order to generate the generalized hypergeometric functions as the normalized function of the coherent states, we can choose the eigenvalues of Hamiltonian operator in two manner.

This method was previously used in the literature [12], [13], [17].

There are *two ways* to define the eigenvalues of the Hamiltonian operator.

1.) We will note the first expression with $e_{BG}(m)$ and call it the *Barut-Girardello (BG) manner* (the justification will be clear in what follows):

$$e_{BG}(m) = m\frac{\prod_{j=1}^{q}\left(b_j + m - 1\right)}{\prod_{i=1}^{p}\left(a_i + m - 1\right)} \quad , \quad m = 1, 2, ..., n \tag{3.24}$$

where the numbers $a_i$ and $b_j$ are real, then the structure constants of GHG-CSs are

$$\rho_{BG}(n) \equiv \prod_{m=1}^{n} e_{BG}(m) = n! \frac{\prod_{j=1}^{q} (b_j)_n}{\prod_{i=1}^{p} (a_i)_n} \ , \ \rho_{BG}(n) = n \frac{\prod_{j=1}^{q} (b_j + m - 1)}{\prod_{i=1}^{p} (a_i + m - 1)} \rho_{BG}(n-1) = e_{BG}(n) \rho_{BG}(n-1) \quad (3.25)$$

that is, just the structure constants that appear in the definition of generalized hypergeometric functions ${}_pF_q(\boldsymbol{a}; \boldsymbol{b}; |z|^2)$.

2.) The second expression, which will turn out to be *dual to the first*, we will write $e_{KP}(m)$ and call it the *Klauder-Perelomov (KP) manner*:

$$e_{KP}(m) = m \frac{\prod_{i=1}^{p} (a_i + m - 1)}{\prod_{j=1}^{q} (b_j + m - 1)} \quad , \quad m = 1, 2, ..., n \qquad (3.26)$$

and similar

$$\rho_{KP}(n) \equiv \prod_{m=1}^{n} e_{KP}(m) = n! \frac{\prod_{i=1}^{p} (a_i)_n}{\prod_{j=1}^{q} (b_j)_n} \ , \ \rho_{KP}(n) = n \frac{\prod_{i=1}^{p} (a_i + m - 1)}{\prod_{j=1}^{q} (b_j + m - 1)} \rho_{KP}(n-1) = e_{KP}(n) \rho_{KP}(n-1) \quad (3.27)$$

which leads to the *dual* normalization function of type ${}_qF_p(\boldsymbol{b}; \boldsymbol{a}; |z|^2)$.

To simplify the calculation of the formulas, we will introduce the notations:

$$B_q(m) \equiv \prod_{j=1}^{q} (b_j + m - 1), \ A_p(m) \equiv \prod_{i=1}^{p} (a_i + m - 1) \ , \ e_{BG}(m) = m \frac{B_q(m)}{A_p(m)} \qquad (3.28)$$

$$\{B_q(n)\} \equiv \prod_{j=1}^{q} (b_j)_n \ , \ \{A_p(n)\} \equiv \prod_{i=1}^{p} (a_i)_n \ , \ \rho_{BG}(n) \equiv n! \frac{\{B_q(n)\}}{\{A_p(n)\}} \qquad (3.29)$$

and similarly for the (KP) manner.

Depending on the type of coherent states we want to use, we will choose the appropriate form of Hamilton's operator, (BG) or (KP). Correspondingly, Hamilton's operators will also be denoted by $\hat{\mathcal{H}}_{BG}$ , their dual by $\hat{\mathcal{H}}_{KP}$, and the eigenvalue equations will be:

$$\begin{aligned} \hat{\mathcal{H}}_{BG} | n \rangle &= \hbar \omega e_{BG}(n) | n \rangle = \hbar \omega n \frac{B_q(n)}{A_p(n)} | n \rangle \quad , \\ \hat{\mathcal{H}}_{KP} | n \rangle &= \hbar \omega e_{KP}(n) | n \rangle = \hbar \omega n \frac{A_p(n)}{B_q(n)} | n \rangle \end{aligned} \qquad (3.30)$$

We will pointed out that there are three fundamental kinds of coherent states: Barut-Girardello, Klauder-Perelomov and Gazeau-Klauder [20]. As Klauder (one of the "parents" of the coherent states formalism) stated "*there are a vast number of coherent state sets, which are distinguished from each other by the presence of different weight factor sets* $\rho(n)$" [21]. This means that the structure functions $\rho(n)$ (and, of course, the energy eigenvalues $e(n)$ that enter into its definition) are the most important entities in the definition of CSs. The definitions of these three kinds of coherent states and their expansions according to the Fock vectors are the following:

1.) ***Coherent states of Barut-Girardello manner (BG-CSs)***, defined as the eigenvectors of annihilation operator $\hat{\mathcal{A}}_-$ [22]:

$$\hat{\mathcal{A}}_- | z >_{BG} = z | z >_{BG} \quad , \quad | z >_{BG} = \frac{1}{\sqrt{{}_pF_q\left(a ;\, b ; | z |^2\right)}} \sum_{n=0}^{\infty} \left[ \frac{\{A_p(n)\}}{\{B_q(n)\}} \right]^{1/2} \frac{z^n}{\sqrt{n!}} | n > \tag{3.31}$$

$$| z >_{BG} = \frac{1}{\sqrt{{}_pF_q\left(a ;\, b ; | z |^2\right)}} \sum_{n=0}^{\infty} \frac{z^n}{\sqrt{\rho_{BG}(n)}} | n > = \frac{1}{\sqrt{{}_pF_q\left(a ;\, b ; | z |^2\right)}} | z))_{BG} \quad , \qquad \rho_{BG}(n) \equiv n! \frac{\{B_q(n)\}}{\{A_p(n)\}} \tag{3.32}$$

2.) ***Coherent states of Klauder-Perelomov manner (KP-CSs)***, defined as the result of the action of the exponential creation operator $\exp\left(z\hat{\mathcal{A}}_+\right)$ (or, in general, of the displacement operator $\exp\left(z\hat{\mathcal{A}}_+ - z^* \hat{\mathcal{A}}_-\right)$) on the vacuum state $|0>$ [23]:

$$| z >_{KP} = \frac{1}{\sqrt{{}_qF_p\left(b ;\, a ; | z |^2\right)}} \exp\left(z\hat{\mathcal{A}}_+\right) | 0 > \tag{3.33}$$

$$| z >_{KP} = \frac{1}{\sqrt{{}_qF_p\left(b ;\, a ; | z |^2\right)}} \sum_{n=0}^{\infty} \frac{z^n}{\sqrt{\rho_{KP}(n)}} | n > = \frac{1}{\sqrt{{}_qF_p\left(b ;\, a ; | z |^2\right)}} | z))_{KP} \quad , \qquad \rho_{KP}(n) \equiv n! \frac{\{A_p(n)\}}{\{B_q(n)\}} = \frac{(n!)^2}{\rho_{BG}(n)} \tag{3.34}$$

3.) ***Coherent states of Gazeau-Klauder manner (GK-CSs)***, defined as a real two-parameter set of coherent states $\{| J , \gamma >_{GK}\}$, with $J > 0$ and $-\infty < \gamma < +\infty$ [20], [24]:

$$| J , \gamma >_{GK} = \frac{1}{\sqrt{N(J)}} \sum_{n=0}^{\infty} \frac{\left(\sqrt{J}\right)^n \exp\left(-\mathrm{i}\gamma e_n\right)}{\sqrt{\rho_{GK}(n)}} | n > \quad , \quad \rho_{GK}(n) = \prod_{m=1}^{n} e_m \tag{3.35}$$

The first two types of coherent states (BG- and KP-) are *dual states* (for details, see [25]). Their non-normalized coherent states are

$$| z))_{BG} = \sum_{n=0}^{\infty} \frac{z^n}{\sqrt{\rho_{BG}(n)}} | n > = \sum_{n=0}^{\infty} \sqrt{\frac{\{A_p(n)\}}{\{B_q(n)\}}} \frac{z^n}{\sqrt{n!}} | n > \tag{3.36}$$

$$| z))_{KP} = \sum_{n=0}^{\infty} \frac{z^n}{\sqrt{\rho_{KP}(n)}} | n > = \sum_{n=0}^{\infty} \sqrt{\frac{\{B_q(n)\}}{\{A_p(n)\}}} \frac{z^n}{\sqrt{n!}} | n > \tag{3.37}$$

The action of the Hamiltonian's operator on the dual non-normalized coherent states are, then

$$\hat{\mathcal{H}}_{BG} \,|\, z))_{BG} = \hbar\,\omega \sum_{n=0}^{\infty} n \frac{B_q(n)}{A_p(n)} \sqrt{\frac{\{A_p(n)\}}{\{B_q(n)\}}} \frac{z^n}{\sqrt{n!}} \,|\, n >= \hbar\,\omega\, e_{BG}\left( z \frac{\partial}{\partial z} \right) | \, z))_{BG} \tag{3.38}$$

$$\hat{\mathcal{H}}_{KP} \,|\, z))_{KP} = \hbar\,\omega \sum_{n=0}^{\infty} n \frac{A_p(n)}{B_q(n)} \sqrt{\frac{\{B_q(n)\}}{\{A_p(n)\}}} \frac{z^n}{\sqrt{n!}} \,|\, n >== \hbar\,\omega\, e_{KP}\left( z \frac{\partial}{\partial z} \right) | \, z))_{KP} \tag{3.39}$$

For two dual kinds of coherent states, the non-normalized matrix elements are

$$((z^* \,|\, \hat{\mathcal{H}}_{BG} \,|\, z'))_{BG} = \hbar\,\omega\, e_{BG}\left( z^* \frac{\partial}{\partial z^*} \right) ((z^* \,|\, z))_{BG} = \hbar\,\omega\, e_{BG}\left( z^* \frac{\partial}{\partial z^*} \right) {}_pF_q\left( \boldsymbol{a};\, \boldsymbol{b};\, z^* z' \right) \tag{3.40}$$

$$((z^* \,|\, \hat{\mathcal{H}}_{GK} \,|\, z'))_{GK} = \hbar\,\omega\, e_{GK}\left( z^* \frac{\partial}{\partial z^*} \right) ((z^* \,|\, z))_{KP} = \hbar\,\omega\, e_{GK}\left( z^* \frac{\partial}{\partial z^*} \right) {}_qF_p\left( \boldsymbol{b};\, \boldsymbol{a};\, z^* z' \right) \tag{3.41}$$

as well as the non-normalized density operator

$$\begin{gathered} ((z^* \,|\, \hat{\Omega}_{BG} \,|\, z'))_{BG} \equiv ((z^* \,|\, e^{-\beta \hat{\mathcal{H}}_{BG}} \,|\, z'))_{BG} = \\ = \sum_{n=0}^{\infty} e^{-\beta\hbar\omega e_{BG}(n)} \frac{\left( z^* z' \right)^n}{\rho_{BG}(n)} = e^{-\beta\hbar\omega e_{BG}\left( z^* \frac{\partial}{\partial z^*} \right)} \, {}_pF_q\left( \boldsymbol{a};\, \boldsymbol{b};\, z^* z' \right) \end{gathered} \tag{3.42}$$

$$\begin{gathered} ((z^* \,|\, \hat{\Omega}_{GK} \,|\, z'))_{GK} \equiv ((z^* \,|\, e^{-\beta \hat{\mathcal{H}}_{KP}} \,|\, z'))_{GK} = \\ == \sum_{n=0}^{\infty} e^{-\beta\hbar\omega e_{KP}(n)} \frac{\left( z^* z' \right)^n}{\rho_{KP}(n)} = e^{-\beta\hbar\omega e_{KP}\left( z^* \frac{\partial}{\partial z^*} \right)} \, {}_qF_p\left( \boldsymbol{b};\, \boldsymbol{a};\, z^* z' \right) \end{gathered} \tag{3.43}$$

Let's calculate the derivative with respect to $\beta$ the matrix elements of the non-normalized density operator taking into account the above results.

For BG-CSs we will have:

$$\begin{gathered} -\frac{\partial}{\partial \beta} ((z^* \,|\, \hat{\Omega}_{BG} \,|\, z'))_{BG} = \left[ -\frac{\partial}{\partial \beta} e^{-\beta\hbar\omega e_{BG}\left( z^* \frac{\partial}{\partial z^*} \right)} \right] {}_pF_q\left( \boldsymbol{a};\, \boldsymbol{b};\, z^* z' \right) = \\ = \hbar\,\omega\, e_{BG}\left( z^* \frac{\partial}{\partial z^*} \right) ((z^* \,|\, \hat{\Omega}_{BG} \,|\, z'))_{BG} = \hat{\mathcal{H}}_{BG}\left( z^* \frac{\partial}{\partial z^*} \right) ((z^* \,|\, \hat{\Omega}_{BG} \,|\, z'))_{BG} \end{gathered} \tag{3.44}$$

so that the Bloch equation becomes

$$-\frac{\partial}{\partial \beta} ((z^* \,|\, \hat{\Omega}_{BG} \,|\, z'))_{BG} = \hat{\mathcal{H}}_{BG}\left( z^* \frac{\partial}{\partial z^*} \right) ((z^* \,|\, \hat{\Omega}_{BG} \,|\, z'))_{BG} \tag{3.45}$$

For KP-CSs the calculations are similar:

$$-\frac{\partial}{\partial\beta}((z^*|\hat{\Omega}_{KP}|z'))_{KP}=\left[-\frac{\partial}{\partial\beta}e^{-\beta\hbar\omega e_{KP}\left(z^*\frac{\partial}{\partial z^*}\right)}\right]{}_qF_p\left(\boldsymbol{b};\,\boldsymbol{a};\,z^*z'\right)=$$
$$=\hbar\omega e_{KP}\left(z^*\frac{\partial}{\partial z^*}\right)((z^*|\hat{\Omega}_{KP}|z'))_{KP}=\hat{\mathcal{H}}_{KP}\left(z^*\frac{\partial}{\partial z^*}\right)((z^*|\hat{\Omega}_{KP}|z'))_{KP} \tag{3.46}$$

so that the Bloch equation becomes

$$-\frac{\partial}{\partial\beta}((z^*|\hat{\Omega}_{KP}|z'))_{KP}=\hat{\mathcal{H}}_{KP}\left(z^*\frac{\partial}{\partial z^*}\right)((z^*|\hat{\Omega}_{KP}|z'))_{KP} \tag{3.47}$$

It can therefore be observed that *the duality of BG-CSs versus KP-CSs is preserved*: from the mathematical point of view the forms of the equations are the same, only are exchanged the indices $p\leftrightarrow q$, respectively the set of numbers $\boldsymbol{a}\leftrightarrow\boldsymbol{b}$ [25].

From here follows a first *rule*: considering the operational equality $z^*\frac{\partial}{\partial z^*}=z^*z'\frac{\partial}{\partial z^*z'}$, in which the variable $z'$ can be regarded as a constant, the matrix elements of Hamilton's operator are obtained by correspondence:

$$((z^*|\hat{\mathcal{H}}_{BG}|z'))_{BG}\rightarrow e_{BG}(n)\quad\Leftrightarrow\quad\hat{\mathcal{H}}_{BG}\left(z^*\frac{\partial}{\partial z^*}\right)\rightarrow e_{BG}\left(z^*\frac{\partial}{\partial z^*}\right)=e_{BG}\left(z^*z'\frac{\partial}{\partial z^*z'}\right)$$
$$((z^*|\hat{\mathcal{H}}_{KP}|z'))\rightarrow e_{KP}(n)\quad\Leftrightarrow\quad\hat{\mathcal{H}}_{KP}\left(z^*\frac{\partial}{\partial z^*}\right)\rightarrow e_{KP}\left(z^*\frac{\partial}{\partial z^*}\right)=e_{KP}\left(z^*z'\frac{\partial}{\partial z^*z'}\right) \tag{3.48}$$

that is, each energy quantum number $n$ in the expression of energy eigenvalue $e(n)$ will be replaced by the operator $z^*\frac{\partial}{\partial z^*}$ (or $z^*z'\frac{\partial}{\partial z^*z'}$, or another operator of this type, e.g. $x\frac{\partial}{\partial x}$, where $x\equiv e^{-\beta\hbar\omega}z^*z'$, on which the generalized hypergeometric function depends). This is, in fact, as we mentioned before, *the main idea of the present paper*.

For the above choice of the dimensionless expression of eigenvalues, the following correspondence is valid:

$$e_{BG}(m)=m\frac{\prod_{j=1}^{q}\left(b_j+m-1\right)}{\prod_{i=1}^{p}\left(a_i+m-1\right)}=m\frac{B_q(n)}{A_p(n)}\;\Leftrightarrow$$
$$\Leftrightarrow\;\hat{\mathcal{H}}_{BG}\left(z^*\frac{\partial}{\partial z^*}\right)=\hbar\omega\frac{\prod_{j=1}^{q}\left(b_j+z^*\frac{\partial}{\partial z^*}-1\right)z^*\frac{\partial}{\partial z^*}}{\prod_{i=1}^{p}\left(a_i+z^*\frac{\partial}{\partial z^*}-1\right)}=\hbar\omega\frac{\left\{\hat{B}_q\left(z^*\frac{\partial}{\partial z^*}\right)\right\}}{\left\{\hat{A}_p\left(z^*\frac{\partial}{\partial z^*}\right)\right\}} \tag{3.49}$$

$$e_{KP}(m) = m\frac{\prod_{i=1}^{p}(a_i + m - 1)}{\prod_{j=1}^{q}(b_j + m - 1)} = m\frac{A_p(n)}{B_q(n)} \Leftrightarrow$$

$$\Leftrightarrow \quad \hat{\mathcal{H}}_{KP}\left(z^*\frac{\partial}{\partial z^*}\right) = \hbar\omega\frac{\prod_{i=1}^{p}\left(a_i + z^*\frac{\partial}{\partial z^*} - 1\right)z^*\frac{\partial}{\partial z^*}}{\prod_{j=1}^{q}\left(b_j + z^*\frac{\partial}{\partial z^*} - 1\right)} = \hbar\omega\frac{\left\{\hat{A}_p\left(z^*\frac{\partial}{\partial z^*}\right)\right\}}{\left\{\hat{B}_q\left(z^*\frac{\partial}{\partial z^*}\right)\right\}} \qquad (3.50)$$

Let's emphasize that the above operator, because the variables $z^*$ and $z'$ appear in the form of a product, can also be expressed as follows:

$$z^*\frac{\partial}{\partial z^*} \equiv z^*z'\frac{\partial}{\partial z^*z'} \equiv e^{-\beta\hbar\omega}z^*z'\frac{\partial}{\partial\left(e^{-\beta\hbar\omega}z^*z'\right)} \qquad (3.51)$$

but for now, due to the lack of calculations, where there is no confusion, we will only use the expression through the variable $z^*$.

If $z = z'$, the diagonal matrix elements of the Hamiltonian in the representation of the *normalized coherent states* are

$$< z^* | \hat{\mathcal{H}}_{BG} | z >_{BG} = \frac{\hbar\omega}{{}_pF_q\left(\boldsymbol{a};\boldsymbol{b};|z|^2\right)}\sum_{n=0}^{\infty}e_{BG}(n)\frac{\left(|z|^2\right)^n}{\rho_{BG}(n)} = \frac{\hbar\omega}{{}_pF_q\left(\boldsymbol{a};\boldsymbol{b};|z|^2\right)}\sum_{n=0}^{\infty}\frac{\left(|z|^2\right)^n}{\rho_{BG}(n-1)} = \hbar\omega|z|^2 \qquad (3.52)$$

where is obtained after renaming the summation index $m = n - 1$ and giving up the non-physical term with $m = -1$.This result is exactly the expression of the relationship called "*action identity*", a requirement imposed on the coherent states of systems with an infinite energy spectrum, within the so-called "*Klauder's prescriptions*" [20].

Similarly, the non-normalized canonical density operator can be written as follows

$$((z^* | \hat{\Omega}_{BG} | z'))_{BG} \equiv ((z^* | e^{-\beta\hat{\mathcal{H}}_{BG}} | z'))_{BG} = \sum_{n=0}^{\infty}e^{-\beta\hbar\omega e_{BG}(n)}\frac{\left(z^*z'\right)^n}{\rho(n)} = e^{-\beta\hbar\omega e_{BG}\left(z^*\frac{\partial}{\partial z^*}\right)}\,{}_pF_q\left(\boldsymbol{a};\boldsymbol{b};z^*z'\right) \qquad (3.53)$$

Also, starting from the relationship

$$\hat{\mathcal{H}}_{BG}\,\hat{\Omega}_{BG}\,|z'))_{BG} = \hat{\mathcal{H}}_{BG}\,\hat{\Omega}_{BG}\sum_{n=0}^{\infty}\frac{(z')^n}{\sqrt{\rho_{BG}(n)}}|n> = \hbar\omega\sum_{n=0}^{\infty}e_{BG}(n)\,e^{-\beta\hbar\omega e(n)}\frac{(z')^n}{\sqrt{\rho_{BG}(n)}}|n> \qquad (3.54)$$

we will get

$$((z^* | \hat{\mathcal{H}}_{BG} \hat{\Omega}_{BG} | z'))_{BG} = \hbar\omega e_{BG}\left( z^* \frac{\partial}{\partial z^*} \right) e^{-\beta\hbar\omega e\left( z^* \frac{\partial}{\partial z^*} \right)} {}_pF_q\left(\boldsymbol{a};\, \boldsymbol{b}; z^* z'\right) =$$
$$= \hbar\omega e_{BG}\left( z^* \frac{\partial}{\partial z^*} \right)((z^* | \hat{\Omega}_{BG} | z'))_{BG} = \hat{\mathcal{H}}_{BG}\left( z^* \frac{\partial}{\partial z^*} \right)((z^* | \hat{\Omega}_{BG} | z'))_{BG} \tag{3.55}$$

so that the Bloch equation becomes

$$-\frac{\partial}{\partial\beta}((z^* | \hat{\Omega}_{BG} | z'))_{BG} = \hat{\mathcal{H}}_{BG}\left( z^* \frac{\partial}{\partial z^*} \right)((z^* | \hat{\Omega}_{BG} | z'))_{BG} \tag{3.56}$$

and similarly for the KP-CSs.

## 4. Role of generalized hypergeometric coherent states

From the relations above, it can be seen that the matrix elements of different physical observables, in the representation of the GHG-CSs, are closely related to the normalized function of the generalized coherent states, i.e. to the functions ${}_pF_q\left(\boldsymbol{a};\, \boldsymbol{b}; z^* z'\right)$ or ${}_qF_p\left(\boldsymbol{b};\, \boldsymbol{a}; z^* z'\right)$ . Thus, the role of this function is crucial in the entire approach to the topic of the canonical density matrix versus canonical Bloch's equation.

Taking advantage of the fact that generalized hypergeometric functions are infinite power series, the matrix elements of the operators in CSs representation can be obtained in another way. It is based on the following property of Pochhammer index: $(x)_n = (x+n-1)(x)_{n-1}$ . For example, for Hamilton's operator in the BG-CSs representation we will have

$$((z^* |\hat{\mathcal{H}}_{BG} |z'))_{BG} = \hbar\omega \sum_n e_{BG}(n) \frac{\left(z^* z'\right)^n}{\rho(n)} =$$
$$= \hbar\omega \sum_n n \frac{\prod_{j=1}^{q}\left(b_j + n - 1\right)}{\prod_{i=1}^{p}\left(a_i + n - 1\right)} \frac{\prod_{i=1}^{p}\left(a_i\right)_n}{\prod_{j=1}^{q}\left(b_j\right)_n} \frac{\left(z^* z'\right)^n}{n!} = \hbar\omega z^* z' \sum_n \frac{\prod_{i=1}^{p}\left(a_i\right)_{n-1}}{\prod_{j=1}^{q}\left(b_j\right)_{n-1}} \frac{\left(z^* z'\right)^{n-1}}{(n-1)!} \tag{4.01}$$

By changing the summation index $m = n-1$ and neglecting the unphysical term with $m = n-1$, we obtain the same result as in accordance with the general case.

$$((z^* |\hat{\mathcal{H}}_{BG} |z'))_{BG} = \hbar\omega\ z^* z' \sum_{m=0}^{\infty} \frac{\prod_{i=1}^{p}\left(a_i\right)_m}{\prod_{j=1}^{q}\left(b_j\right)_m} \frac{\left(z^* z'\right)^m}{m!} = \hbar\omega\ z^* z'\ {}_pF_q\left(\boldsymbol{a};\, \boldsymbol{b};\ z^* z'\right) \tag{4.02}$$

Of course, the same procedure can be applied to the left-hand side of the Bloch equation

$$-\frac{\partial}{\partial\beta}((z^*|\hat{\Omega}_{BG}|z'))_{BG} = -\frac{\partial}{\partial\beta}\sum_{n=0}^{\infty} e^{-\beta\hbar\omega e_{BG}(n)}\frac{(z^*z')^n}{\rho(n)} = \hbar\omega\sum_{n=0}^{\infty} e_{BG}(n)e^{-\beta\hbar\omega e_{BG}(n)}\frac{(z^*z')^n}{\rho(n)} =$$
$$= \hbar\omega e^{-\beta\hbar\omega e_{BG}\left(z^*\frac{\partial}{\partial z^*}\right)}\sum_n n\frac{B_q(n)}{A_p(n)}\frac{(z^*z')^n}{\rho(n)} = \hbar\omega z^*z' e^{-\beta\hbar\omega e_{BG}\left(z^*\frac{\partial}{\partial z^*}\right)}{}_pF_q(a;\, b;\, z^*z') = \tag{4.03}$$
$$= \hbar\omega z^*z'((z^*|\hat{\Omega}_{BG}|z'))_{BG}$$

having in mind that the operators $e_{BG}\left(z^*\frac{\partial}{\partial z^*}\right)$ and $e^{-\beta\hbar\omega e_{BG}\left(z^*\frac{\partial}{\partial z^*}\right)}$ are commutable.

Similar relations are obtained in the case of the representation of KP-CSs.

In conclusion, for any type of coherent states BG-CSs or KP-CSs, the Bloch equation in the representation of coherent states has the expression:

$$-\frac{\partial}{\partial\beta}((z^*|\hat{\Omega}(\beta)|z')) = \hat{\mathcal{H}}\left(z^*\frac{\partial}{\partial z^*}\right)((z^*|\hat{\Omega}(\beta)|z')) \tag{4.04}$$

In the right-hand side of the equation appears the Hamiltonian operator of the system $\hat{\mathcal{H}}\left(z^*\frac{\partial}{\partial z^*}\right)$ in the GHG-CSs representation, which is constructed so that it contains the $z^*\frac{\partial}{\partial z^*}$ operator, by replacing each principal quantum number $n$ of energy eigenvalues expression $e(n)$ with the $z^*\frac{\partial}{\partial z^*}$ operator. So we have the following correspondence between energy eigenvalues versus the Hamiltonian operator:

$$n \to z^*\frac{\partial}{\partial z^*} \quad\Leftrightarrow\quad \hbar\omega e(n) \to \hat{\mathcal{H}}\left(z^*\frac{\partial}{\partial z^*}\right) \quad\Leftrightarrow\quad e(n) \to e\left(z^*\frac{\partial}{\partial z^*}\right) \tag{4.05}$$

We emphasize once again that Hamilton's operator $\hat{\mathcal{H}}\left(z^*\frac{\partial}{\partial z^*}\right)$ acts only on the variable on the left $z^*$, which belongs to the space of bra vectors, $((z^*|$.

In particular, the particle number operator $\hat{\mathcal{N}}|n> = n|n>$, acting on a non-normalized coherent state leads to the expression:

$$\hat{\mathcal{N}}|z')) = \sum_n\frac{(z')^n}{\sqrt{\rho(n)}}\hat{\mathcal{N}}|n> = \sum_n\frac{(z')^n}{\sqrt{\rho(n)}}n\,|n> = z'\frac{\partial}{\partial z'}|z')) \quad , \qquad ((z^*|\hat{\mathcal{N}} = z^*\frac{\partial}{\partial z^*}((z^*| \tag{4.06}$$

which additionally justifies the correspondence $n \to \hat{\mathcal{N}} \to z^*\frac{\partial}{\partial z^*}$ (or, equivalently

$n \to \hat{\mathcal{H}} \to z^* z' \frac{\partial}{\partial z^* z'}$ ), because $\hat{\mathcal{H}} = \hat{\mathcal{H}}\left(\hat{\mathcal{N}}\right)$.

Proceeding similarly as above, the action of Hamilton's operator on the canonical density matrix is then (see, Appendix A1).

$$\hat{\mathcal{H}}\left( z^* \frac{\partial}{\partial z^*} \right)((z^* \mid \hat{\Omega}(\beta) \mid z')) = \hbar \omega z^* z'((z^* \mid \hat{\Omega}(\beta) \mid z')) \tag{4.07}$$

This means that generally for quantum systems with energy spectrum $E(n) = \hbar \omega e(n)$ the Bloch equation will be written

$$-\frac{\partial}{\partial \beta}((z^* \mid \hat{\Omega}(\beta) \mid z') = \hbar \omega z^* z'((z^* \mid \hat{\Omega}(\beta) \mid z') \tag{4.08}$$

If we integrate this equation

$$\int_{\Omega(0)}^{\Omega(\beta)} \frac{d((z^* \mid \hat{\Omega}(\beta) \mid z'))}{((z^* \mid \hat{\Omega}(\beta) \mid z'))} = -\hbar \omega z^* z' \int_0^{\beta} d\beta \tag{4.09}$$

we obtain a possible particular solution

$$((z^* \mid \hat{\Omega}(\beta) \mid z')) = ((z^* \mid \hat{\Omega}(0) \mid z')) \exp\left(-\beta \hbar \omega \, z^* z'\right) \tag{4.10}$$

This solution satisfies the boundary condition

$$\lim_{\beta \to 0} ((z^* \mid \hat{\Omega}(\beta) \mid z')) = ((z^* \mid \hat{\Omega}(0) \mid z')) = {}_pF_q\left(\boldsymbol{a};\, \boldsymbol{b};\, z^* z'\right) \tag{4.11}$$

Finally, the particular solution of the Bloch equation will be

$$((z^* \mid \hat{\Omega}(\beta) \mid z')) = {}_pF_q\left(\boldsymbol{a};\, \boldsymbol{b};\, z^* z'\right) \exp\left(-\beta \hbar \omega \, z^* z'\right) \tag{4.12}$$

As two illustrative examples, in the next section we will calculate the expression of the density matrix for the case of quantum systems that have a linear, respectively quadratic energy spectrum.

To simplify the writing of formulas, a new variable will be used: $x \equiv e^{-\beta \hbar \omega} z^* z'$, and $z^* \frac{\partial}{\partial z^*} = x \frac{\partial}{\partial x}$, so the action of the Hamiltonian's operator on the generalized hypergeometric function is

$$\hat{\mathcal{H}}\left( x \frac{\partial}{\partial x} \right) {}_pF_q(\boldsymbol{a};\, \boldsymbol{b}\,;\, x) = \hbar \omega \, \frac{\hat{B}\left( x \frac{\partial}{\partial x} \right)}{\hat{A}\left( x \frac{\partial}{\partial x} \right)} {}_pF_q(\boldsymbol{a};\, \boldsymbol{b}\,;x) = \hbar \omega \; x \; {}_pF_q(\boldsymbol{a};\, \boldsymbol{b}\,;x\,) \tag{4.13}$$

This equation is of the type of eigenvalue equations: for the Hamiltonian operator $\hat{\mathcal{H}}\left(x\frac{\partial}{\partial x}\right)$, the eigenfunctions are ${}_pF_q(\boldsymbol{a};\, \boldsymbol{b}\,;\, x)$, and the eigenvalues are $\hbar\,\omega\, x$.

The last equality can be formally written as

$$\hat{B}\left(x\frac{\partial}{\partial x}\right){}_pF_q(\boldsymbol{a};\, \boldsymbol{b}\,;\, x) = x\,\hat{A}\left(x\frac{\partial}{\partial x}\right)\,{}_pF_q(\boldsymbol{a};\, \boldsymbol{b}\,;\, x) \tag{4.14}$$

Taking into account the definition of operators $\hat{A}$ and $\hat{B}$, we will obtain a new differential equation that is satisfied by the generalized hypergeometric functions:

$$\left[\prod_{j=1}^{q}\left(x\frac{\partial}{\partial x}+b_j-1\right)x\frac{\partial}{\partial x}-x\prod_{i=1}^{p}\left(x\frac{\partial}{\partial x}+a_i-1\right)\right]{}_pF_q(\boldsymbol{a};\, \boldsymbol{b}\,; x)=0 \tag{4.15}$$

This new differential equation that we obtained is different from the one that appears, as a rule, in the literature (see, for example, [26], [27]), due to the different expression of Hamiltonian's operator in the coherent states representation, as the *ratio* of operators $\hat{A}\left(x\frac{\partial}{\partial x}\right)$ and $\hat{B}\left(x\frac{\partial}{\partial x}\right)$.

The correctness of the equation obtained by us can be demonstrated by another method, for example by using the integral representation of the generalized geometric function [15] (see, Appendix A2):

$${}_pF_q(\boldsymbol{a};\, \boldsymbol{b}\,;\, x) = \frac{\prod_{j=1}^{q}\Gamma(b_j)}{\prod_{i=1}^{p}\Gamma(a_i)}\frac{1}{2\pi\mathrm{i}}\int_L \frac{\prod_{i=1}^{p}\Gamma(a_i+s)}{\prod_{j=1}^{q}\Gamma(b_j+s)}\Gamma(-s)(-x)^s\, ds \tag{4.16}$$

where we suppose that $L$ is a contour that starts at infinity on a line parallel to the positive real axis, encircles the nonnegative integers in the negative sense, and ends at infinity on another line parallel to the positive real axis. The contour of integration separates the poles of $\Gamma(a_i+s),\ i=1,2,...,p$, from those of $\Gamma(-s)$.

The same method, applied to the KP-CSs representation, leads to perfectly similar results, taking into account of course the duality of the previously mentioned indices and sets of numbers.

**a) The case of systems with linear spectrum**

For the *linear quantum systems* the energy eigenvalue is

$$E(n) = \hbar\,\omega e(n) \equiv \hbar\,\omega(n+e_0) \tag{4.17}$$

For this case $p=q=1$ and also $\boldsymbol{a}=1;\ \boldsymbol{b}=e_0+1$, as well as the corresponding generalized hypergeometric function is

$$ {}_1F_1\left(1;\, e_0+1;\, z^*z'\right)=\sum_{n=0}^{\infty}\frac{(1)_n}{(e_0+1)_n}\frac{(z^*z')^n}{n!} \tag{4.18}$$

and also the following correspondence

$$ \hat{\mathcal{H}}_{BG}\left(z^*\frac{\partial}{\partial z^*}\right)\to \hbar\omega e_{BG}\left(z^*\frac{\partial}{\partial z^*}\right)=\hbar\omega\left(z^*\frac{\partial}{\partial z^*}+e_0\right) \tag{4.19}$$

The corresponding canonical density operator is

$$ \hat{\Omega}^{(l)}(\beta)=e^{-\beta\hbar\omega e_0}\sum_{n=0}^{\infty}\left(e^{-\beta\hbar\omega}\right)^n |\, n><n\,| \tag{4.20}$$

and the non-normalized density matrix in the representation of coherent states becomes

$$ ((z^*\,|\,\hat{\Omega}^{(L)}(\beta)|\,z'))_{BG}=e^{-\beta\hbar\omega e_0}\sum_{n=0}^{\infty}\frac{(1)_n}{(e_0+1)_n}\frac{\left(e^{-\beta\hbar\omega}z^*z'\right)^n}{n!}=e^{-\beta\hbar\omega e_0}\,{}_1F_1\left(1;\, e_0+1\,;e^{-\beta\hbar\omega}z^*z'\right) \tag{4.21}$$

So, from the point of view of the calculations, they do not differ from the previous ones, with the only observation that the argument of the generalized hypergeometric function is now multiplied by the factor $\exp(-\beta\hbar\omega)$. That is why we can write the result directly:

$$ \begin{aligned} \hat{\mathcal{H}}_{BG}\left(z^*\frac{\partial}{\partial z^*}\right)((z^*\,|\,\hat{\Omega}^{(L)}(\beta)|\,z'))_{BG} &=\hbar\omega e^{-\beta\hbar\omega e_0}\left(z^*\frac{\partial}{\partial z^*}+e_0\right){}_1F_1\left(1;\, e_0+1\,;e^{-\beta\hbar\omega}z^*z'\right)= \\ &=\hbar\omega\left(z^*\frac{\partial}{\partial z^*}+e_0\right)((z^*\,|\,\hat{\Omega}^{(L)}(\beta)|\,z'))_{BG} \end{aligned} \tag{4.22}$$

Consequently, for the quantum systems with linear energy spectra $E(n)=\hbar\omega(n+e_0)$, the Bloch's equation has the form

$$ -\frac{\partial}{\partial\beta}((z^*\,|\,\hat{\Omega}^{(L)}(\beta)|\,z'))_{BG}=\hbar\omega\left(z^*\frac{\partial}{\partial z^*}+e_0\right)((z^*\,|\,\hat{\Omega}^{(L)}(\beta)|\,z'))_{BG} \tag{4.23}$$

The correctness of this equation, that is, the solution (4.21), can be checked if the following operational relations are used:

$$ \begin{gathered} x\equiv e^{-\beta\hbar\omega}z^*z', \\ -\frac{\partial}{\partial\beta}=-\frac{\partial}{\partial e^{-\beta\hbar\omega}}\frac{\partial e^{-\beta\hbar\omega}}{\partial\beta}=\hbar\omega e^{-\beta\hbar\omega}\frac{\partial}{\partial e^{-\beta\hbar\omega}}=\hbar\omega x\frac{\partial}{\partial x}\quad,\quad z^*\frac{\partial}{\partial z^*}=x\frac{\partial}{\partial x} \end{gathered} \tag{4.24}$$

After the corresponding derivations, it is found that the two members of Bloch's equation are equal:

$$-\frac{\partial}{\partial\beta}((z^*|\hat{\Omega}^{(L)}(\beta)|z'))_{BG} = -\frac{\partial}{\partial\beta}\left[e^{-\beta\hbar\omega e_0}\sum_{n=0}^{\infty}\frac{\left(e^{-\beta\hbar\omega}z^*z'\right)^n}{\rho(n)}\right] =$$

$$= \hbar\omega e_0((z^*|\hat{\Omega}^{(L)}(\beta)|z'))_{BG} + \hbar\omega\, e^{-\beta\hbar\omega e_0} e^{-\beta\hbar\omega} z^*z'\sum_{m=0}^{\infty}\frac{\left(e^{-\beta\hbar\omega}z^*z'\right)^m}{\rho(m)} = \tag{4.25}$$

$$= \hbar\omega\left(e_0 + e^{-\beta\hbar\omega}z^*z'\right)((z^*|\hat{\Omega}^{(L)}(\beta)|z'))_{BG}$$

$$\hbar\omega\left(z^*\frac{\partial}{\partial z^*} + e_0\right)((z^*|\hat{\Omega}^{(L)}(\beta)|z'))_{BG} = \hbar\omega\left(z^*\frac{\partial}{\partial z^*} + e_0\right)\left[e^{-\beta\hbar\omega e_0}\sum_{n=0}^{\infty}\frac{\left(e^{-\beta\hbar\omega}z^*z'\right)^n}{\rho(n)}\right] =$$

$$= \hbar\omega e^{-\beta\hbar\omega e_0} e^{-\beta\hbar\omega} z^*z'\sum_{m=0}^{\infty}\frac{\left(e^{-\beta\hbar\omega}z^*z'\right)^m}{\rho(m)} + \hbar\omega e_0((z^*|\hat{\Omega}^{(L)}(\beta)|z'))_{BG} = \tag{4.26}$$

$$= \hbar\omega\left(e_0 + e^{-\beta\hbar\omega}z^*z'\right)((z^*|\hat{\Omega}^{(L)}(\beta)|z'))_{BG}$$

where in both relations we have eliminated the non-physical term with $m = -1$.

Consequently, the Bloch equation for systems with linear energy spectra becomes

$$-\frac{\partial}{\partial\beta}((z^*|\hat{\Omega}^{(L)}(\beta)|z'))_{BG} = \hbar\omega\left(e_0 + e^{-\beta\hbar\omega}z^*z'\right)((z^*|\hat{\Omega}^{(L)}(\beta)|z'))_{BG} \tag{4.27}$$

Integrating the equation between $0$ and $\beta$, we get

$$((z^*|\hat{\Omega}^{(L)}(\beta)|z'))_{BG} = ((z^*|\hat{\Omega}^{(L)}(0)|z'))_{BG}\exp\left[\left(e^{-\beta\hbar\omega} - 1\right)z^*z' - \beta\hbar\omega e_0\right] =$$
$$= {}_1F_1\left(1;\, e_0 + 1;\, z^*z'\right)\exp\left[\left(e^{-\beta\hbar\omega} - 1\right)z^*z' - \beta\hbar\omega e_0\right] \tag{4.28}$$

$$\lim_{\beta\to 0}((z^*|\hat{\Omega}^{(L)}(\beta)|z'))_{BG} = ((z^*|\hat{\Omega}^{(L)}(0)|z'))_{BG} = {}_1F_1\left(1;\, e_0 + 1;\, z^*z'\right) \tag{4.29}$$

On the other hand, equating the two ways of expressing the non-normalized density matrix (the expression from the definition and the one obtained as a solution to Bloch's equation), we will have:

$${}_1F_1\left(1;\, e_0 + 1;\, e^{-\beta\hbar\omega}z^*z'\right) = {}_1F_1\left(1;\, e_0 + 1;\, z^*z'\right)\exp\left[\left(e^{-\beta\hbar\omega} - 1\right)z^*z'\right] \tag{4.30}$$

Taking into account the relationships

$${}_2F_1(a, b;\, a;\, x) = {}_1F_0(b;\, ;\, x) = \frac{1}{(1-x)^b} \quad , \quad {}_0F_0(\, ;\, ;\varsigma) = e^x \tag{4.31}$$

as well as the relation that expresses the product of two generalized hypergeometric functions, we will be able to verify the correctness of the relation resulting from the integration of Bloch's equation [28].

$$
{}_pF_q\left(\boldsymbol{a};\boldsymbol{b};x\right){}_rF_s\left(\boldsymbol{c};\boldsymbol{d};g\,x\right)=
$$

$$
=\sum_{m=0}^{\infty}\frac{\prod_{i=1}^{p}(a_i)_m}{\prod_{j=1}^{q}(b_j)_m}\,{}_{q+r+1}F_{p+s}\left(-m,1-m-\boldsymbol{b},\boldsymbol{c};1-m-\boldsymbol{a};\boldsymbol{d};(-1)^{p+q+1}g\right)\frac{x^m}{m!} \tag{4.32}
$$

We will have, successively

$$
{}_1F_1\left(1;e_0+1;z^*z'\right){}_0F_0\left(;;\left(e^{-\beta\hbar\omega}-1\right)z^*z'\right)=
$$

$$
=\sum_{m=0}^{\infty}\frac{(1)_m}{(e_0+1)_m}\,{}_2F_1\left(-m,1-m-e_0-1;1-m-1;-\left(e^{-\beta\hbar\omega}-1\right)\right)\frac{\left(z^*z'\right)^m}{m!}=
$$

$$
=\sum_{m=0}^{\infty}\frac{(1)_m}{(e_0+1)_m}\,{}_1F_0\left(-m-e_0;;-\left(e^{-\beta\hbar\omega}-1\right)\right)\frac{\left(z^*z'\right)^m}{m!}=e^{-\beta\hbar\omega e_0}\sum_{m=0}^{\infty}\frac{(1)_m}{(e_0+1)_m}\frac{\left(e^{-\beta\hbar\omega}z^*z'\right)^m}{m!}= \tag{4.33}
$$

$$
=e^{-\beta\hbar\omega e_0}\,{}_1F_1\left(1;e_0+1;e^{-\beta\hbar\omega}\,z^*z'\right)=((z^*\,|\,\hat{\Omega}^{(l)}(\beta)\,|\,z'))_{BG}
$$

The ratio of the two generalized hypergeometric functions is

$$
\frac{{}_1F_1\left(1;e_0+1;e^{-\beta\hbar\omega}z^*z'\right)}{{}_1F_1\left(1;e_0+1;z^*z'\right)}=\exp\left[\left(e^{-\beta\hbar\omega}-1\right)z^*z'\right] \tag{4.34}
$$

Reminding us that, generally, the Husimi's distribution function is obtained as an expected value of the *normalized* density operator based on coherent states $(z=z')$, as [2]

$$
Q^{(H)}\left(|z|^2\right)\equiv<z^*\,|\,\rho\,|\,z>_{BG}=\frac{1}{Z(\beta)}\frac{1}{{}_pF_q\left(\boldsymbol{a};\boldsymbol{b};|z|^2\right)}\sum_n e^{-\beta E(n)}\frac{\left(|z|^2\right)^n}{\rho(n)} \tag{4.35}
$$

For linear energy systems, the non-normalized Husimi’s function becomes:

$$
\begin{aligned}
Q^{(H)}_{\text{non-norm}}\left(|z|^2\right)&\equiv((z\,|\,\rho\,|\,z))=e^{-\beta\hbar\omega e_0}\frac{{}_1F_1\left(1;e_0+1;e^{-\beta\hbar\omega}\,|z|^2\right)}{{}_1F_1\left(1;e_0+1;|z|^2\right)}=\\
&=e^{-\beta\hbar\omega e_0}\exp\left[\left(e^{-\beta\hbar\omega}-1\right)|z|^2\right]
\end{aligned} \tag{4.36}
$$

For large values of arguments, the confluent hypergeometric function ${}_1F_1(a;b;x)$ can be approximated as [30]

$$
{}_1F_1(a;b;x)\approx\frac{\Gamma(b)}{\Gamma(a)}e^x\,x^{a-b} \tag{4.37}
$$

so that for large values of $e^{-\beta\hbar\omega}z^*z'$ the asymptotic solution becomes

$$((z^* \,|\, \hat{\Omega}^{(L)}(\beta)|\, z'))_{BG} \approx \Gamma(e_0+1)e^{-\beta\hbar\omega e_0}\frac{\exp\left(e^{-\beta\hbar\omega} z^* z'\right)}{\left(z^* z'\right)^{e_0}} \tag{4.38}$$

We previously stated that by customizing the indices $p$ and $q$ of generalized hypergeometric function, all the normalization functions can be obtained, respectively all the coherent states associated with them. Let's illustrate the above with some examples related to some simple quantum systems - quantum oscillators.

**4.1 One Dimensional Harmonic Oscillator (HO-1D).** For HO-1D all three kinds (definitions) of coherent states (BG-CS, KP-CSs and GK-CSs are equivalent.

$$e(n) = n + \frac{1}{2} \quad \Leftrightarrow \quad \hat{\mathcal{H}}\left(z^* \frac{\partial}{\partial z^*}\right) = \hbar\omega\left(z^* \frac{\partial}{\partial z^*} + \frac{1}{2}\right) \;, \quad e_0 = \frac{1}{2} \tag{4.39}$$

The indices $p$ and $q$ of the generalized hypergeometric function are $p=1$ , $q=1$ , and $a_1 = 1$ , $b_1 = 3/2$, the structure function is $\rho(n) = (3/2)_n$ The canonical density matrix is

$$\begin{aligned} ((z^* \,|\, \hat{\Omega}^{(HO)}(\beta)|\, z')) &= \sum_{n=0}^{\infty} e^{-\beta\hbar\omega e(n)} \frac{(1)_n}{(3/2)_n} \frac{\left(z^* z'\right)^n}{n!} = e^{-\beta\frac{\hbar\omega}{2}} \sum_{n=0}^{\infty} \frac{(1)_n}{(3/2)_n} \frac{\left(e^{-\beta\hbar\omega} z^* z'\right)^n}{n!} = \\ &= e^{-\beta\frac{\hbar\omega}{2}} {}_1F_1\left(1\,;\,3/2;\,e^{-\beta\hbar\omega} z^* z'\right) = \frac{\sqrt{\pi}}{2} \frac{\exp\left(e^{-\beta\hbar\omega} z^* z'\right)}{\sqrt{e^{-\beta\hbar\omega} z^* z'}} \operatorname{erf}\left(\sqrt{e^{-\beta\hbar\omega} z^* z'}\right) \end{aligned} \tag{4.40}$$

To avoid this relatively complicated expression, in practical calculations the zero energy is given up, $e_0 = 0$ , so that $e(n) = n$ , $\rho(n) = n!$ and

$$((z^* \,|\, \hat{\Omega}^{(HO)}(\beta)|\, z')) = \sum_{n=0}^{\infty} e^{-\beta\hbar\omega n} \frac{\left(z^* z'\right)^n}{n!} = \exp\left(e^{-\beta\hbar\omega} z^* z'\right) \tag{4.41}$$

The differential equation with partial derivatives (the canonical Bloch equation) is

$$-\frac{\partial}{\partial\beta}((z^* \,|\, \hat{\Omega}^{(HO)}(\beta)|\, z')) = \hbar\omega z^* \frac{\partial}{\partial z^*}((z^* \,|\, \hat{\Omega}^{(HO)}(\beta)|\, z')) \tag{4.42}$$

Because the canonical matrix of the density is symmetric in the variables $z^*$ and $z'$, we will look for the solution in the form

$$((z^* \,|\, \hat{\Omega}^{(HO)}(\beta)|\, z')) = \exp\left[f(\beta) z^* z'\right] \tag{4.43}$$

The solution is easily obtained, and after integration between $0$ and $\beta$, this is

$$((z^* \,|\, \hat{\Omega}^{(HO)}(\beta)|\, z')) = ((z^* \,|\, \hat{\Omega}^{(HO)}(0)|\, z'))\exp\left(e^{-\beta\hbar\omega} z^* z'\right) \tag{4.44}$$

At the $\beta \to 0$ limit we will get the boundary condition:

$$\lim_{\beta\to 0}((z^* \mid \hat{\Omega}(\beta) \mid z')) = e^{z^* z'} = {}_0F_0\left(/;/;z^* z'\right) \tag{4.45}$$

Then, the corresponding *normalized* density matrix is [30]

$$\begin{aligned} < z^* \mid \hat{\Omega}^{(HO)}(\beta) \mid z' > &= \frac{1}{\sqrt{{}_0F_0\left(/;/;\mid z\mid^2\right)}} \frac{1}{\sqrt{{}_0F_0\left(/;/;\mid z'\mid^2\right)}} ((z^* \mid \hat{\Omega}(\beta) \mid z')) = \\ &= e^{-\beta\frac{\hbar\omega}{2}} \exp\left[ -\frac{1}{2}\left(\mid z\mid^2 + \mid z'\mid^2\right) + e^{-\beta\hbar\omega} z^* z' \right] \end{aligned} \tag{4.46}$$

**4.2 Pseudoharmonic Oscillator (PHO).** For PHO, the three types (definitions) of coherent states (Barut-Girardello, Klauder-Perelomov and Gazeau-Klauder) are not equivalent, they lead to different expressions. That is why we will examine them separately.

a) ***Barut-Girardello coherent states (BG-CSs).***

The structure function of these states, as well as the Hamiltonian's correspondence are

$$\begin{gathered} \rho_{BG}^{(PHO)}(n) = \prod_{m=1}^{n} e_{BG}^{(PHO)}(m) = (k+1)_n \quad \Leftrightarrow \quad e_{BG}^{(PHO)}(m) = m + k \\ \hat{\mathcal{H}}_{BG}^{(PHO)}\left( z^* \frac{\partial}{\partial z^*} \right) = \hbar\omega \left( z^* \frac{\partial}{\partial z^*} + k \right) \end{gathered} \tag{4.47}$$

where $k$ is the Bargmann index labeling the irreducible representations of the corresponding quantum group (its values can be natural integers or half-integers) corresponds to the normalization function with $p = 1$ and $q = 1$, with $a_1 = 1$ , $b_1 = k + 1$

$${}_1F_1\left(1;k+1;z^* z'\right) = \sum_{n=0}^{\infty} \frac{(1)_n}{(k+1)_n} \frac{\left(z^* z'\right)^n}{n!} \tag{4.48}$$

Consequently, the BG-CSs of the pseudoharmonic oscillator are

$$\mid z,k >_{BG}^{(PHO)} = \frac{1}{\sqrt{{}_1F_1\left(1;k+1;\mid z\mid^2\right)}} \sum_{n=0}^{\infty} \sqrt{\frac{(1)_n}{(k+1)_n}} \frac{z^n}{\sqrt{n!}} \mid n,k > \tag{4.49}$$

The non-normalized density matrix of PHO, in the BG-CSs representation is

$$\begin{aligned} ((z^*, k \mid \hat{\Omega}_{BG}^{(PHO)}(\beta) \mid z', k)) &= e^{-\beta\hbar\omega k} \sum_{n=0}^{\infty} \frac{(1)_n}{(k+1)_n} \frac{\left(e^{-\beta\hbar\omega} z^* z'\right)^n}{n!} = \\ &= e^{-\beta\hbar\omega k} {}_1F_1\left(1;k+1;e^{-\beta\hbar\omega} z^* z'\right) \end{aligned} \tag{4.50}$$

and the canonical Bloch equation for non-normalized density matrix $((z^*, k \mid \hat{\Omega}_{BG}^{(PHO)}(\beta) \mid z', k))$ is

$$-\frac{\partial}{\partial\beta} ((z^*, k \mid \hat{\Omega}_{BG}^{(PHO)}(\beta) \mid z', k)) = \hbar\omega \left( z^* \frac{\partial}{\partial z^*} + k \right) ((z^*, k \mid \hat{\Omega}_{BG}^{(PHO)}(\beta) \mid z', k)) \tag{4.51}$$

Let's look for a solution like that

$$((z^*,k\,|\,\hat{\Omega}_{BG}^{(PHO)}(\beta)|\,z',k)) = e^{-\beta\hbar\omega k}\,{}_1F_1\left(1;k+1;e^{-\beta\hbar\omega}z^*z'\right) \tag{4.52}$$

with the boundary condition

$$\lim_{\beta\to 0}((z^*,k\,|\,\hat{\Omega}_{BG}^{(PHO)}(\beta)|\,z',k)) = ((z^*,k\,|\,\hat{\Omega}_{BG}^{(PHO)}(0)|\,z',k)) = {}_1F_1\left(1;k+1;z^*z'\right) \tag{4.53}$$

**b) *Klauder-Perelomov coherent states (KP-CSs)***

These states are dual versus the BG-CSs. The structure function of KP-CSs of PHO, as well as the Hamiltonian's correspondence are

$$e_{KP}^{(PHO)}(m) = \frac{m}{2k+m-1} \quad \Leftrightarrow \quad \rho_{KP}^{(PHO)}(n) = \prod_{m=1}^{n} e_{KP}^{(PHO)}(m) = \frac{n!}{(2k)_n} \quad \Leftrightarrow$$
$$\hat{\mathcal{H}}_{KP}\left(z^*\frac{\partial}{\partial z^*}\right) = \hbar\,\omega\left(z^*\frac{\partial}{\partial z^*} + 2k\right) \tag{4.54}$$

The normalization function with $p=0$ and $q=1$, with $b_1 = 2k+1$

$${}_1F_0\left(2k;\ ;z^*z'\right) = \sum_{n=0}^{\infty}(2k)_n\frac{\left(z^*z'\right)^n}{n!} = \frac{1}{\left(1-z^*z'\right)^{2k}} \tag{4.55}$$

Consequently, the Klauder-Perelomov coherent states of the PHO are [31]

$$|\,z,k>_{KP}^{(PHO)} = \left(1-|\,z\,|^2\right)^k\sum_{n=0}^{\infty}\sqrt{(2k)_n}\,\frac{z^n}{\sqrt{n!}}|n,k> \tag{4.56}$$

The non-normalized density matrix of PHO, in the KP-CSs representation is

$$\begin{aligned}((z^*,k\,|\,\hat{\Omega}_{KP}^{(PHO)}(\beta)|\,z',k)) &= e^{-\beta\hbar\omega\,2k}\sum_{n=0}^{\infty}(2k)_n\frac{\left(e^{-\beta\hbar\omega}z^*z'\right)^n}{n!} = \\ &= e^{-\beta\hbar\omega\,2k}\,{}_1F_0\left(2k;\ ;e^{-\beta\hbar\omega}\,z^*z'\right) = e^{-\beta\hbar\omega\,2k}\frac{1}{\left(1-e^{-\beta\hbar\omega}\,z^*z'\right)^{2k}}\end{aligned} \tag{4.57}$$

The corresponding canonical Bloch equation is, then

$$-\frac{\partial}{\partial\beta}((z^*,k\,|\,\hat{\Omega}_{KP}^{(PHO)}(\beta)|\,z',k)) = \hbar\,\omega\left(z^*\frac{\partial}{\partial z^*} + 2k\right)((z^*,k\,|\,\hat{\Omega}_{KP}^{(PHO)}(\beta)|\,z',k)) \tag{4.58}$$

The solution that verifies this equation, including the boundary condition, is

$$((z^*,k\,|\,\hat{\Omega}_{KP}^{(PHO)}(\beta)|\,z',k)) = e^{-\beta\hbar\omega\,2k}\,{}_1F_0\left(2k;\ ;e^{-\beta\hbar\omega}\,z^*z'\right) = e^{-\beta\hbar\omega\,2k}\frac{1}{\left(1-e^{-\beta\hbar\omega}\,z^*z'\right)^{2k}} \tag{4.59}$$

$$\lim_{\beta\to 0}((z^*,k\,|\,\hat{\Omega}_{KP}^{(PHO)}(\beta)|\,z',k))={}_1F_0\left(2k;\;;z^*z'\right)=\frac{1}{\left(1-z^*z'\right)^{2k}} \tag{4.60}$$

**c) *Gazeau-Klauder coherent states (GK-CSs).*** By using the substitution $\omega_{PHO}\equiv\omega=2\omega_0$ (the corresponding one-dimensional harmonic oscillator HO-1D has the angular frequency $\omega_0$), the structure function of these states, as well as the Hamiltonian's correspondence are

$$e_{GK}^{(PHO)}(m)=2(m+k),\quad \rho_{GK}^{(PHO)}(n)=\prod_{m=1}^{n}e_{GK}^{(PHO)}(m)=2^n\,(k+1)_n$$
$$\hat{\mathcal{H}}_{GK}\left(z^*\frac{\partial}{\partial z^*}\right)\;\rightarrow\;\hbar\,\omega\left(\sqrt{J}\,\frac{\partial}{\partial\sqrt{J}}+k\right) \tag{4.61}$$

The normalization function with $p=1$ and $q=1$, with $a_1=1$ , $b_1=k+1$ is

$${}_1F_1\left(1\,;k+1;z^*z'\right)=\sum_{n=0}^{\infty}\frac{(1)_n}{(k+1)_n}\frac{\left(z^*z'\right)^n}{n!} \tag{4.62}$$

Then, the Gazeau-Klauder coherent states (GK-CSs) of the pseudoharmonic oscillator are indexed by replacing the complex variable $z$ by two independent real numbers $J$ and $\gamma$, so that $J\geq 0$ and $-\infty<\gamma<+\infty$, namely $z=\sqrt{J}\exp(-i\gamma)$. In this way, the new obtained coherent states was denoted by $|\,J\,,\gamma>$. Their expressions are [32]

$$|\,J\,,\gamma>_{GK}^{(PHO)}=\frac{1}{\sqrt{{}_1F_1\left(1\,;k+1;|\,z\,|^2\right)}}\sum_{n=0}^{\infty}\sqrt{\frac{(1)_n}{(k+1)_n}}\frac{\left(\sqrt{\frac{J}{2}}\right)^n}{\sqrt{n!}}|n,k> \tag{4.63}$$

The non-normalized density matrix of PHO, in the GK-CSs representation is

$$((J,\gamma\,|\,\hat{\Omega}_{GK}^{(PHO)}(\beta)|\,J\,',\gamma))=e^{-\beta\hbar\omega k}\sum_{n=0}^{\infty}\frac{(1)_n}{(k+1)_n}\frac{\left(e^{-\beta\hbar\omega}\,\frac{\sqrt{J\,J\,'}}{2}\right)^n}{n!}=$$
$$=e^{-\beta\hbar\omega k}\;{}_1F_1\left(1;k+1;e^{-\beta\hbar\omega}\,\frac{\sqrt{J\,J\,'}}{2}\right) \tag{4.64}$$

At the same time, the canonical Bloch equation for non-normalized density matrix is

$$-\frac{\partial}{\partial\beta}((J,\gamma\,|\,\hat{\Omega}_{GK}^{(PHO)}(\beta)|\,J\,',\gamma))=\hbar\,\omega\left(\sqrt{J}\,\frac{\partial}{\partial\sqrt{J}}+k\right)((J,\gamma\,|\,\hat{\Omega}_{GK}^{(PHO)}(\beta)|\,J\,',\gamma)) \tag{4.65}$$

Rearranging the right member, we will have (with $x \equiv e^{-\beta\hbar\omega}\frac{\sqrt{JJ'}}{2}$), let's consider the following operational equalities:

$$-\frac{\partial}{\partial\beta}=\hbar\omega e^{-\beta\hbar\omega}\frac{\partial}{\partial e^{-\beta\hbar\omega}}=\hbar\omega\left(e^{-\beta\hbar\omega}\frac{\sqrt{JJ'}}{2}\right)\frac{\partial}{\partial\left(e^{-\beta\hbar\omega}\frac{\sqrt{JJ'}}{2}\right)}=\hbar\omega x\frac{\partial}{\partial x} \tag{4.66}$$

and also

$$\sqrt{J}\frac{\partial}{\partial\sqrt{J}}=\left(e^{-\beta\hbar\omega}\frac{\sqrt{JJ'}}{2}\right)\frac{\partial}{\partial\left(e^{-\beta\hbar\omega}\frac{\sqrt{JJ'}}{2}\right)}=x\frac{\partial}{\partial x} \tag{4.67}$$

we will have in the left member of the equation:

$$\begin{aligned} &-\frac{\partial}{\partial\beta}((J,\gamma\,|\,\hat{\Omega}_{GK}^{(PHO)}(\beta)|\,J',\gamma))= \\ &=\hbar\omega k\,((J,\gamma\,|\,\hat{\Omega}_{GK}^{(PHO)}(\beta)|\,J',\gamma))-e^{-\beta\hbar\omega k}\frac{\partial}{\partial\beta}\,{}_1F_1(1;k+1;x)= \\ &=\hbar\omega k\,((J,\gamma\,|\,\hat{\Omega}_{GK}^{(PHO)}(\beta)|\,J',\gamma))+\hbar\omega e^{-\beta\hbar\omega k}\,x\frac{\partial}{\partial x}{}_1F_1(1;k+1;x)\end{aligned} \tag{4.68}$$

and the same, in the right member:

$$\begin{aligned} &\hbar\omega\left(J\frac{\partial}{\partial J}+k\right)((J,\gamma\,|\,\hat{\Omega}_{GK}^{(PHO)}(\beta)|\,J',\gamma))=\hbar\omega\left(\frac{1}{2}\varsigma\frac{\partial}{\partial\varsigma}+k\right)((J,\gamma\,|\,\hat{\Omega}_{GK}^{(PHO)}(\beta)|\,J',\gamma))= \\ &=\hbar\omega e^{-\beta\hbar\omega k}\,x\frac{\partial}{\partial x}\,{}_1F_1(1;k+1;x)+\hbar\omega k\,((J,\gamma\,|\,\hat{\Omega}_{GK}^{(PHO)}(\beta)|\,J',\gamma>\end{aligned} \tag{4.69}$$

Without performing the derivation of the generalized hypergeometric function, it is observed that the two members of the equation are equal.

This shows that the expression (4.64) for the non-normalized density matrix is indeed a solution of the Bloch equation, and its boundary condition is fulfilled:

$$\lim_{\beta\to 0}((J,\gamma\,|\,\hat{\Omega}_{GK}^{(PHO)}(\beta)|\,J',\gamma))=((J,\gamma\,|\,\hat{\Omega}_{GK}^{(PHO)}(0)|\,J',\gamma))={}_1F_1\left(1;k+1;\frac{\sqrt{JJ'}}{2}\right) \tag{4.70}$$

**b) The case of quadratic spectrum systems**. For *non-linear systems,* which have a *quadratic energy spectrum* the situation is more complicated, in the sense that a specific ansatz must be used (introduced for the first time for the Morse oscillator, see [33]).

The dimensionless energy eigenvalues for the quadratic energy spectrum and $a_1 = 1$ , $b_1 = 1$ , $b_2 = b+1$, are

$$e(n) = n(n+b) \ , \quad b \neq 0 \quad , \quad \rho(n) = \prod_{m=1}^{n} e(n) = n!(b+1)_n \tag{4.71}$$

The Hamiltonian's operator is, after substitution $n \Leftrightarrow z^* z' \frac{\partial}{\partial z^* z'}$ :

$$\hat{\mathcal{H}}\left( z^* \frac{\partial}{\partial z^*} \right) = \hbar \omega z^* \frac{\partial}{\partial z^*} \left( z^* \frac{\partial}{\partial z^*} + b \right) \tag{4.72}$$

From the dimensionless expression of energy, we will make the following notations:

$$\beta E(n) \equiv \beta \hbar \omega n(n+b) = \beta \hbar \omega n^2 + \beta \hbar \omega b n \equiv \varepsilon n^2 + \varepsilon b \, n, \quad \varepsilon \equiv \beta \hbar \omega \tag{4.78}$$

The canonical density matrix (non- normalized) in coherent states representation for these kind of systems is, according the definition

$$((z^* \,|\, \hat{\Omega}^{(q)}(\beta) \,|\, z')) = \sum_{n=0}^{\infty} e^{-\varepsilon n(n+b)} \frac{1}{(b+1)_n} \frac{(z^* z')^n}{n!} =$$

$$= \sum_{n=0}^{\infty} e^{-\varepsilon n^2} \frac{1}{(b+1)_n} \frac{(e^{-\varepsilon b} z^* z')^n}{n!} = e^{-\varepsilon \left( z^* \frac{\partial}{\partial z^*} \right)^2} {}_0F_1\left( \ ; \ b+1; \, e^{-\varepsilon b} z^* z' \right) \tag{4.79}$$

where the index $(q)$ means that it is the non-normalized density matrix of a system with a quadratic energy spectrum.

Our *ansatz* consists in developing in power series the quadratic term in energy exponential $\exp[-\beta E(n)]$, so that we will have, successively:

$$e^{-\beta E(n)} = e^{-\varepsilon n^2} \left( e^{-\varepsilon b} \right)^n = \sum_{m=0}^{\infty} \frac{(-\varepsilon)^m}{m!} n^{2m} \left( e^{-\varepsilon b} \right)^n =$$

$$= \sum_{m=0}^{\infty} \frac{(-\varepsilon)^m}{m!} \left( \frac{\partial}{\partial \varepsilon b} \right)^{2m} \left( e^{-\varepsilon b} \right)^n = \exp\left[ -\frac{1}{b^2} \varepsilon \left( \frac{\partial}{\partial \varepsilon} \right)^2 \right] \left( e^{-\varepsilon b} \right)^n \tag{4.80}$$

Then, the elements of the non-normalized density matrix in the representation of coherent states are

$$((z^* \,|\, \hat{\Omega}^{(q)}(\beta) \,|\, z')) = \exp\left[ -\frac{1}{b^2} \varepsilon \left( \frac{\partial}{\partial \varepsilon} \right)^2 \right] \sum_{n=0}^{\infty} \frac{1}{(b+1)_n} \frac{(e^{-\varepsilon b} z^* z')^n}{n!} =$$

$$= \exp\left[ -\frac{1}{b^2} \varepsilon \left( \frac{\partial}{\partial \varepsilon} \right)^2 \right] {}_0F_1\left( \ ; \ b+1; \, e^{-\varepsilon b} z^* z' \right) \tag{4.81}$$

From these last relations, the following equivalence relation between the operators can be revealed:

$$\exp\left[-\varepsilon\left(z^*\frac{\partial}{\partial z^*}\right)^2\right]=\exp\left[-\frac{1}{b^2}\varepsilon\left(\frac{\partial}{\partial\varepsilon}\right)^2\right] \tag{4.82}$$

Let's derive in relation to beta the definition relation of the matrix elements of the canonical matrix:

$$\begin{aligned}
-\frac{\partial}{\partial\beta}((z^*\,|\,\hat{\Omega}^{(q)}(\beta)|\,z')) &= -\hbar\,\omega\,\frac{\partial}{\partial\varepsilon}((z^*\,|\,\hat{\Omega}^{(q)}(\beta)|\,z'))=\\
&=\hbar\,\omega\sum_{n=0}^{\infty}n\,(n+b)\,e^{-\varepsilon\,n\,(n+b)}\,\frac{1}{(b+1)_n}\,\frac{\left(e^{-\varepsilon b}z^*z'\right)^n}{n!}=\\
&=\hbar\,\omega\,z^*\frac{\partial}{\partial z^*}\left(z^*\frac{\partial}{\partial z^*}+b\right)e^{-\varepsilon\left(z^*\frac{\partial}{\partial z^*}\right)^2}\sum_{n=0}^{\infty}\frac{1}{(b+1)_n}\,\frac{\left(e^{-\varepsilon b}z^*z'\right)^n}{n!}=\\
&=\hbar\,\omega\,z^*\frac{\partial}{\partial z^*}\left(z^*\frac{\partial}{\partial z^*}+b\right)e^{-\varepsilon\left(z^*\frac{\partial}{\partial z^*}\right)^2}{}_0F_1\left(\;;\;b+1;\;e^{-\varepsilon b}z^*z'\right)
\end{aligned} \tag{4.83}$$

The matrix elements of Hamilton's operator applied to $((z^*\,|\,\hat{\Omega}^{(q)}(\beta)|\,z'))$ leads to the same result:

$$\hat{\mathcal{H}}\left(z^*\frac{\partial}{\partial z^*}\right)((z^*\,|\,\hat{\Omega}^{(q)}(\beta)|\,z'))=\hbar\,\omega\,z^*\frac{\partial}{\partial z^*}\left(z^*\frac{\partial}{\partial z^*}+b\right)e^{-\varepsilon\left(z^*\frac{\partial}{\partial z^*}\right)^2}{}_0F_1\left(\;;\;b+1;\;e^{-\varepsilon b}z^*z'\right) \tag{4.84}$$

Therefore, the $((z^*\,|\,\hat{\Omega}^{(q)}(\beta)|\,z'))$ satisfies the canonical Bloch equation

$$-\frac{\partial}{\partial\beta}((z^*\,|\,\hat{\Omega}^{(q)}(\beta)|\,z'))=\hat{\mathcal{H}}\left(z^*\frac{\partial}{\partial z^*}\right)((z^*\,|\,\hat{\Omega}^{(q)}(\beta)|\,z')) \tag{4.85}$$

or, written in the following form:

$$-\frac{\partial}{\partial\beta}((z^*\,|\,\hat{\Omega}^{(q)}(\beta)|\,z'))=\hbar\,\omega\,z^*\frac{\partial}{\partial z^*}\left(z^*\frac{\partial}{\partial z^*}+b\right)((z^*\,|\,\hat{\Omega}^{(q)}(\beta)|\,z')) \tag{4.86}$$

Implicitly, this calculation shows that the expression of the matrix elements of the density operator, obtained by its definition, is the correct solution of Bloch's canonical equation. In this way, the often complicated solution of the Bloch equation, an equation with partial derivatives of the second order, is avoided.

## 5. Concluding remarks

In this paper, we tried to show some properties of the non-normalized density matrix $\hat{\Omega}(\beta)=\exp\left(-\beta\hat{\mathcal{H}}\right)$, sometimes called the *canonical density matrix* due to the similarity of the corresponding density operator with the one corresponding to the canonical distribution.

The expression of the canonical density matrix can be obtained by two methods. The first, and the simplest, is to calculate the expression of the density matrix starting from the definition of the density operator in a certain basis (for example, the basis of the Fock vectors) and using a certain representation (the representation of the coordinate, of the momentum or the representation of coherent states). Among these representations, it seems that the representation of generalized coherent states is the easiest to achieve. In this representation, the canonical density matrix $((z^* | \hat{\Omega}(\beta) | z'))$ is proportional with a generalized hypergeometric function ${}_pF_q\left(\boldsymbol{a};\boldsymbol{b};z^*z'\right)$.

For greater clarity, in the paper we highlighted by writing, in the case of the vectors bra $((...|$, the variable in its complex form, for example $((z^*|$. Obviously, in the case of vectors ket $|...))$ the non-conjugate variable appears $|z'))$.

The second method is a bit more complicated. This involves solving Bloch's canonical equation. This equation is a differential equation with partial derivatives, which is often not very easy to solve.

In the paper we overcame this difficulty in the following way: we first constructed the expression of the canonical density matrix in the representation of coherent states (let's denote it, generically, with DM-CSs), knowing of course the expression of the energy eigenvalues of the system. We then demonstrated that Hamilton's operator, in the representation of coherent states, is constructed in such a way that each quantum number $n$ in the expression of the energy eigenvalues is replaced by the operator $z^*\frac{\partial}{\partial z^*}$. The idea comes from the property of the operator's algebra in the complex Segal-Bargmann space, in which the algebra of ladder operators $\hat{\boldsymbol{a}}^+$ and $\hat{\boldsymbol{a}}$ is represented directly by the algebra of the complex variable $z$ and its derivative $\frac{\partial}{\partial z}$, respectively the particle number $n$ by the product $\hat{\boldsymbol{a}}^+\boldsymbol{a}=z\frac{\partial}{\partial z}$. We then wrote Bloch's canonical equation in the representation of coherent states, using the expression of Hamilton's operator obtained according to the method described above. We checked, then, if DM-CSs satisfies Bloch's canonical equation. Where it was simpler, for confirmation, we solved this equation, obviously obtaining the same expression for DM-CSs. For more complicated cases, we also used the differential equation of the generalized hypergeometric function ${}_pF_q\left(\boldsymbol{a};\boldsymbol{b};z^*z'\right)$.

However, the main purpose of the paper was to demonstrate that, in order to use the canonical density matrix, it is possible to avoid solving the canonical Bloch equation, which, as we mentioned before, is a differential equation with partial derivatives.

To support the theoretical considerations presented above, we examined some examples of quantum systems with a linear spectrum (the harmonic oscillator and the pseudoharmonic oscillator, the last one with the three different types of coherent states), as well as systems with a more complicated energy spectrum (quadratic).

**Appendix A 1**

The action of Hamilton's operator on the canonical density matrix is

$$\mathcal{H}\left(z^*\frac{\partial}{\partial z^*}\right)((z^*\,|\,\hat{\Omega}(\beta)|\,z')) = \hbar\,\omega\,\frac{\prod_{j=1}^{q}\left(b_j + z^*\frac{\partial}{\partial z^*} - 1\right)z^*\frac{\partial}{\partial z^*}}{\prod_{i=1}^{p}\left(a_i + z^*\frac{\partial}{\partial z^*} - 1\right)}((z^*\,|\,\hat{\Omega}(\beta)|\,z')) \equiv$$

$$\equiv \hbar\,\omega\frac{\hat{B}\left(z^*\frac{\partial}{\partial z^*}\right)}{\hat{A}\left(z^*\frac{\partial}{\partial z^*}\right)}\sum_{n=0}^{\infty}e^{-\beta\hbar\omega\,e(n)}\frac{\prod_{i=1}^{p}(a_i)_n}{\prod_{j=1}^{q}(b_j)_n}\frac{\left(z^*z'\right)^n}{n!} = \hbar\,\omega\sum_{n=0}^{\infty}e^{-\beta\hbar\omega\,e(n)}\frac{\prod_{i=1}^{p}(a_i)_n}{\prod_{j=1}^{q}(b_j)_n}e\left(z^*\frac{\partial}{\partial z^*}\right)\frac{\left(z^*z'\right)^n}{n!} \tag{A.11}$$

Let us emphasize that the logic of this notation is to highlight that operator $\hat{B}$ refers to the products containing the quantities $b_j$, respectively operator $\hat{A}$ only to the products containing the quantities $a_i$. As a result, we will have:

$$\left(b_j + z^*\frac{\partial}{\partial z^*} - 1\right)z^*\frac{\partial}{\partial z^*}\left(z^*\right)^n = n\left(b_j + n - 1\right)\left(z^*\right)^n \quad , \quad \hat{B}\left(z^*\frac{\partial}{\partial z^*}\right)\left(z^*\right)^n = n\prod_{j=1}^{q}\left(b_j + n - 1\right)\left(z^*\right)^n \tag{A.12}$$

$$\left(a_i + z^*\frac{\partial}{\partial z^*} - 1\right)\left(z^*\right)^n = \left(a_i + n - 1\right)\left(z^*\right)^n \quad , \quad \hat{A}\left(z^*\frac{\partial}{\partial z^*}\right)\left(z^*\right)^n = \prod_{i=1}^{p}\left(a_i + n - 1\right)\left(z^*\right)^n \tag{A.13}$$

$$\frac{\hat{B}\left(z^*\frac{\partial}{\partial z^*}\right)}{\hat{A}\left(z^*\frac{\partial}{\partial z^*}\right)}((z^*\,|\,\hat{\Omega}(\beta)|\,z')) = \frac{\hat{B}\left(z^*\frac{\partial}{\partial z^*}\right)}{\hat{A}\left(z^*\frac{\partial}{\partial z^*}\right)}\sum_{n=0}^{\infty}e^{-\beta\hbar\omega\,e(n)}\frac{\prod_{i=1}^{p}(a_i)_n}{\prod_{j=1}^{q}(b_j)_n}\frac{\left(z^*z'\right)^n}{n!} =$$

$$= e^{-\beta\hbar\omega e\left(z^*\frac{\partial}{\partial z^*}\right)}\sum_{n=0}^{\infty}\frac{\prod_{i=1}^{p}(a_i)_n}{\prod_{j=1}^{q}(b_j)_n}\frac{n\prod_{j=1}^{q}\left(b_j + n - 1\right)}{\prod_{i=1}^{p}\left(a_i + n - 1\right)}\frac{\left(z^*z'\right)^n}{n!} = z^*z'\,e^{-\beta\hbar\omega e\left(z^*\frac{\partial}{\partial z^*}\right)}\sum_{n=0}^{\infty}\frac{\prod_{i=1}^{p}(a_i)_{n-1}}{\prod_{j=1}^{q}(b_j)_{n-1}}\frac{\left(z^*z'\right)^{n-1}}{(n-1)!} \tag{A.14}$$

After changing the summation index $m=n-1$ and eliminating the non-physical term $m=-1$, we can reintroduce the exponential again under the sum sign and the above expression becomes

$$\hat{\mathcal{H}}\left(z^*\frac{\partial}{\partial z^*}\right)((z^*\,|\,\hat{\Omega}(\beta)|\,z')) = \hbar\omega z^* z' \sum_{m=0}^{\infty} e^{-\beta\hbar\omega\, e(m)} \frac{\prod_{i=1}^{p}(a_i)_m}{\prod_{j=1}^{q}(b_j)_m} \frac{(z^* z')^m}{m!} = \tag{A.15}$$

$$= \hbar\omega z^* z' ((z^*\,|\,\hat{\Omega}(\beta)|\,z'))$$

**Appendix A 2**

Applying to the integral representation of generalized hypergeometric function the ratio of operator functions $\hat{B}/\hat{A}$, we will successively obtain

$$\frac{\hat{B}\left(x\frac{\partial}{\partial x}\right)}{\hat{A}\left(x\frac{\partial}{\partial x}\right)}\,{}_pF_q(\boldsymbol{a};\boldsymbol{b};x) = \frac{\prod_{j=1}^{q}\Gamma(b_j)}{\prod_{i=1}^{p}\Gamma(a_i)}\frac{1}{2\pi\mathrm{i}}\int_L \frac{\prod_{i=1}^{p}\Gamma(a_i+s)}{\prod_{j=1}^{q}\Gamma(b_j+s)}\Gamma(-s)\frac{\hat{B}\left(x\frac{\partial}{\partial x}\right)}{\hat{A}\left(x\frac{\partial}{\partial x}\right)}(-x)^s ds =$$

$$= \frac{\prod_{j=1}^{q}\Gamma(b_j)}{\prod_{i=1}^{p}\Gamma(a_i)}\frac{1}{2\pi\mathrm{i}}\int_L \frac{\prod_{i=1}^{p}\Gamma(a_i+s)}{\prod_{j=1}^{q}\Gamma(b_j+s)}\Gamma(-s)(-1)(-s)\frac{\sum_{j=1}^{q}(b_j+s-1)}{\sum_{i=1}^{p}(a_i+s-1)}(-x)^s ds =$$

$$= x\frac{\prod_{j=1}^{q}\Gamma(b_j)}{\prod_{i=1}^{p}\Gamma(a_i)}\frac{1}{2\pi\mathrm{i}}\int_L \frac{\prod_{i=1}^{p}\Gamma(a_i+s-1)\sum_{i=1}^{p}\Gamma(a_i+s-1)}{\prod_{j=1}^{q}\Gamma(b_j+s-1)\sum_{j=1}^{q}\Gamma(b_j+s-1)}\Gamma(-s+1)(-x)^{s-1}d(s-1) = \tag{A.21}$$

$$= x\frac{\prod_{j=1}^{q}\Gamma(b_j)}{\prod_{i=1}^{p}\Gamma(a_i)}\frac{1}{2\pi\mathrm{i}}\int_{-\mathrm{i}\omega}^{+\mathrm{i}\omega} \frac{\prod_{i=1}^{p}\Gamma(a_i+\xi)}{\prod_{j=1}^{q}\Gamma(b_j+\xi)}\Gamma(-\xi)(-x)^{\xi}d\xi = x\,{}_pF_q(\boldsymbol{a};\boldsymbol{b};x)$$

where finally we used a new integration variable $\xi = s-1$.

Therefore, we have

$$\frac{\hat{B}\left(x\frac{\partial}{\partial x}\right)}{\hat{A}\left(x\frac{\partial}{\partial x}\right)}\,{}_pF_q(\boldsymbol{a};\boldsymbol{b};x) = x\,{}_pF_q(\boldsymbol{a};\boldsymbol{b};x) \tag{A.22}$$

such that the differential equation for the generalized hypergeometric function will be:

$$\left[\hat{B}\left(x\frac{\partial}{\partial x}\right)-x\hat{A}\left(x\frac{\partial}{\partial x}\right)\right]{}_pF_q(\boldsymbol{a};\boldsymbol{b};x)=0 \qquad \text{(A.23)}$$

which proves that it is correct. Implicitly, by this we have demonstrated that the canonical density matrix $((z^* \,|\, \hat{\Omega}(\beta) \,|\, z'))$ attached to a quantum system with a linear spectrum is proportional to a generalized hypergeometric function ${}_pF_q(\boldsymbol{a};\boldsymbol{b};x)$.